\documentclass[nonacm]{acmart}
\makeatletter
\let\@authorsaddresses\@empty
\makeatother

\usepackage{tikz}
\usetikzlibrary{shadows}
\usetikzlibrary{matrix}
\usepackage{booktabs}
\usepackage{graphicx}
\usepackage[table]{xcolor}

\begin{document}

\title[Landscape of LLM Value Preferences]
{Growth First, Care Second?\\Tracing the Landscape of LLM Value Preferences in Everyday Dilemmas}

\author{Zhiyi Chen}
\email{zchen346@usc.edu}
\orcid{0009-0009-3060-6277}
\affiliation{%
  \institution{University of Southern California}
  \city{Los Angeles}
  \state{CA}
  \country{USA}}
\affiliation{%
  \institution{Tsinghua University}
  \city{Haidian}
  \state{Beijing}
  \country{China}}

\author{Eun Cheol Choi}
\email{euncheol@usc.edu}
\orcid{0000-0003-0861-1343}
\affiliation{%
  \institution{University of Southern California}
  \city{Los Angeles}
  \state{CA}
  \country{USA}}

\author{Yingjia Luo}
\email{yingjial@usc.edu}

\author{Xinyi Wang}
\email{xwang597@usc.edu}
\affiliation{%
  \institution{University of Southern California}
  \city{Los Angeles}
  \state{CA}
  \country{USA}}
\affiliation{%
  \institution{Tsinghua University}
  \city{Haidian}
  \state{Beijing}
  \country{China}}

\author{Yulei Xiao}
\email{yuleixia@usc.edu}

\author{Aizi Yang}
\email{aiziyang@usc.edu}

\author{Luca Luceri}
\email{lluceri@isi.edu}
\orcid{0000-0001-5267-7484}
\affiliation{%
  \institution{University of Southern California}
  \city{Los Angeles}
  \state{CA}
  \country{USA}}

\renewcommand{\shortauthors}{Chen et al.}

\begin{abstract}
    People increasingly seek advice online from both human peers and large language model (LLM)–based chatbots. Such advice rarely involves identifying a single correct answer; instead, it typically requires navigating trade-offs among competing values. We aim to characterize how LLMs navigate value trade-offs across different advice-seeking contexts. First, we examine the value trade-off structure underlying advice seeking using a curated dataset from four advice-oriented subreddits. Using a bottom-up approach, we inductively construct a hierarchical value framework by aggregating fine-grained values extracted from individual advice options into higher-level value categories. We construct value co-occurrence networks to characterize how values co-occur within dilemmas and find substantial heterogeneity in value trade-off structures across advice-seeking contexts: a women-focused subreddit exhibits the highest network density, indicating more complex value conflicts; women’s, men’s, and friendship-related subreddits exhibit highly correlated value-conflict patterns centered on \textit{security}-related tensions (\textit{security vs.\ respect/connection/commitment}); by contrast, career advice forms a distinct structure where \textit{security} frequently clashes with \textit{self-actualization} and \textit{growth}.
    We then evaluate LLM value preferences against these dilemmas and find that, across models and contexts, LLMs consistently prioritize values related to \textit{Exploration \& Growth} over \textit{Benevolence \& Connection}. This systemically skewed value orientation highlights a potential risk of value homogenization in AI-mediated advice, raising concerns about how such systems may shape decision-making and normative outcomes at scale.
\end{abstract}

\keywords{large language models, advice seeking, value preferences, value homogenization, hierarchical value framework, online communities}

\maketitle

\section{Introduction}

Large language models (LLMs) are becoming increasingly natural, fluent, and socially intuitive in their interactions with users. As these models grow more conversational, users have begun to rely on them not only for information retrieval, but also for advice across personal, social, and professional domains~\cite{cheong2024not,schneiders2025objection,khurana2024and}. While some users engage with LLM advice critically and analytically, others appear to trust or rely on it too readily, indicating a growing dependence on AI-mediated guidance in everyday decision-making~\cite{wester2024exploring,montag2025role}.

Importantly, LLMs are value-laden, and these values are reflected in the advice and recommendations they provide. Existing work finds that LLMs show identifiable value preferences and moral orientations~\cite{han2025dual}. These value expressions may arise from different stages of model development: training data can reproduce societal biases and cultural stereotypes~\cite{yu2023large}, whereas Reinforcement Learning from Human Feedback (RLHF) may introduce subjective value priors from model trainers~\cite{kirk2023past}. Consequently, when LLMs provide advice, they convey values that may influence how users perceive choices, evaluate trade-offs, and make decisions. Understanding such embedded values is therefore critical for characterizing the sociotechnical role of LLMs in advice-seeking.

In light of this, \textit{capturing the value preferences of LLMs} becomes an urgent and critical question. Existing studies rely on the assumption that values can be inferred from the choices models make when confronted with dilemmas~\cite{rawls1951outline, jiang2025investigating}. While this provides a useful starting point, several limitations remain in existing work. First, evaluation benchmarks often rely on GPT-generated scenarios built from predefined templates or hand-crafted designs~\cite{DBLP:conf/iclr/ChiuJ025,sorensen2024value}, constraining scalability and limiting the diversity of dilemmas constructed. Second, such synthetic benchmarks risk overlooking how value trade-offs manifest in real-world settings, and may therefore fail to capture how LLMs handle real-world dilemmas.
Third, analyses of value trade-offs typically rely on top-down, predefined value taxonomies, which may leave many real-world cases unrepresented or underspecified.

To address these limitations, we draw on real-world dilemmas collected from online advice-seeking communities. Platforms such as Reddit contain abundant, organically occurring dilemmas, offering a more ecologically valid view of how individuals negotiate value conflicts~\cite{bhatia2025computational}. Building on this, we first identify the values motivating each option in the dilemmas and construct a hierarchical value framework using a bottom-up inductive approach. We then use this framework to (i) uncover the value trade-off structures underlying advice-seeking in different subreddits, and (ii) assess the value preferences exhibited by several state-of-the-art LLMs when confronted with real-world dilemmas. Based on this, we propose our two research questions:
\begin{itemize}
    \item \textbf{RQ1:} Which values frequently co-occur, and how do such co-occurrences vary across subreddits?
    \item \textbf{RQ2:} What are the overall value preferences exhibited by LLMs when responding to dilemmas discussed on Reddit, and to what extent do such preferences differ across models and contexts?
\end{itemize}

\paragraph{Contributions of This Work}
In this study, we leverage a curated dataset comprising 5,728 real-world dilemmas sourced from four advice-oriented subreddits: r/AskMenAdvice, r/AskWomenAdvice, r/CareerAdvice, and r/FriendshipAdvice. Using GPT-4o, we extract the values motivating each option in the dilemmas, yielding 2,288 unique values in total. We then inductively construct a four-level hierarchical value framework from these values. Based on this framework, we uncover the value trade-off structures that characterize different subreddits. Finally, we present the same real-world dilemmas to three large language models (GPT-4o, DeepSeek-V3.2-Exp, and Gemini-2.5-Flash) and assess their value preferences through the choices they make. The contributions of this work are summarized as follows:
\begin{itemize}
    \item \textbf{Methodological pipeline.} We develop a scalable pipeline for leveraging real-world dilemmas to assess LLMs’ value preferences.
    \item \textbf{Value framework.} We inductively construct a four-level hierarchical value framework from social media data, offering a data-driven alternative to top-down value taxonomies.
    \item \textbf{Empirical findings.} We evaluate several state-of-the-art LLMs on real-world dilemmas and find that across models and contexts, LLMs consistently prioritize \textit{Exploration \& Growth} over \textit{Benevolence \& Connection}, suggesting potential tendencies to privilege self-development over relational care. 
    \item \textbf{Dataset.} We publicly release a dataset of real-world dilemmas containing annotated values organized within our hierarchical value framework, along with LLM choices, to support reproducible analysis of value-related behaviors in LLMs.\footnote{https://github.com/Renesmeeczy/Value-Trade-off-in-Reddit-Dilemmas}
\end{itemize}

\section{Related Work}

\subsection{Everyday Advice-Seeking and Value Trade-offs in Online Communities}

Recent computational social science research shows that everyday advice-seeking on online platforms is best understood as navigating value trade-offs rather than resolving isolated moral questions. Large-scale analyses of Reddit have demonstrated that real-world dilemmas embed multiple competing considerations and that these trade-offs exhibit systematic structure rather than random co-occurrence \cite{bhatia2025computational, nguyen2022mapping}. Related work applying moral and value frameworks to Reddit discussions further shows that value distributions vary across communities, topics, and social contexts \cite{gamage2023moral, park2024valuescope}, indicating that advice forums may also encode distinct normative environments rather than a single moral logic. Together, these findings motivate treating advice-oriented subreddits as sites where value conflicts are repeatedly articulated and negotiated.

Research on relationship and gendered advice communities further suggests that advice is shaped by power dynamics and social expectations. Studies of dating and relationship advice on platforms such as Douban and Instagram show that advice discourse often prioritizes safety, care, and self-protection, particularly for women, reflecting broader structural inequalities \cite{tan2025our, large2025new}. Above works inform RQ1 by suggesting that value trade-off structures may differ systematically across advice communities rather than reflecting a universal pattern.

\subsection{Human Value Frameworks and Computational Representations}

Classical theories of human values treat values as structured, multidimensional systems rather than independent dimensions, laying a foundation for computational modeling~\cite{schwartz1992universals, haidt2004intuitive}. Building on this insight, recent work has shown that values can be inferred from arguments and behavioral traces in large-scale text corpora. Argument-level datasets and stance prediction studies demonstrate that value categories can be reliably extracted from natural language and linked to downstream behaviors \cite{mirzakhmedova2024touche23, zhang2024enhancing}. These approaches establish the feasibility of mapping textual content to value representations at scale.

Another strain of research moves beyond predefined taxonomies toward bottom-up, data-driven value discovery \cite{jiang2025investigating}. Studies extract implicit norms and values from social media interactions or human–AI conversations using embedding-based clustering, with interpretive labels applied post hoc \cite{park2024valuescope, huang2025values}. Large-scale analyses of Reddit dilemmas similarly embed explanations or attributes to reveal recurring value tensions without relying exclusively on hand-crafted categories \cite{bhatia2025computational}. At the same time, work on machine moral judgment cautions that bottom-up approaches can reproduce cultural and social biases present in training data, underscoring the need for validation and stability checks \cite{jiang2025investigating}. This literature motivates our bottom-up hierarchical framework while also shaping its validation strategy.

\subsection{LLMs' Value Preferences, and Homogenization Risks}

A growing literature treats large language models as moral and advice-giving agents whose value preferences can be inferred from their decisions in dilemma-based tasks \cite{jiang2025investigating}. This is especially important as people judge moral justifications and advice from LLMs as trustworthy, maybe even more so than those sourced from ethicists \cite{dillion2025ai}. Benchmark efforts such as \texttt{DailyDilemmas}, \texttt{ValueBench}, and \texttt{Value Compass Benchmarks} evaluate LLMs across multiple moral and value theories, revealing consistent biases and model-specific patterns in value preferences \cite{DBLP:conf/iclr/ChiuJ025, ren2024valuebench, yao2025value}. Complementary work adopts psychological paradigms directly, treating LLM outputs as individual responses and analyzing them using established tools such as Moral Foundations Theory \cite{nunes2024large}. These studies collectively demonstrate that LLMs express structured value preferences rather than neutral or random judgments.

At the same time, recent research raises concerns about the societal risks of value alignment and convergence. Studies show that aligning models to specific value distributions can increase toxicity or bias, and that emotionally adaptive or value-aligned systems may foster dependency, distorted judgment, or harmful reinforcement \cite{choi2025unintended, chu2025illusions}. Critical scholarship cautions that AI systems are not value-neutral, but reflect underlying power asymmetries in data, design, and governance, raising concerns that LLMs may reinforce dominant normative frameworks while marginalizing alternative value systems, which is referred to as \textit{value homogenization} \cite{kamran2023decolonizing, farina2025ethical}. LLMs may consistently amplify certain values while suppressing others, leading to value homogenization that differentially aligns with some users’ preferences while alienating others \cite{shahid2025llms}. Cultural values from underrepresented, low-resource languages may not be well-represented in LLMs, leading to value convergence and uneven performance at scale \cite{zhang2025ethosgpt}. Lack of human diversity in LLMs call for more robust evaluation pipelines and mitigation strategies \cite{anthis2025llm}. These findings directly motivate RQ2 by highlighting the importance of examining whether LLMs converge on uniform value preferences across diverse advice contexts.

\section{Methodology}
\subsection{Data}
We leverage an existing dataset comprising real-world dilemmas collected from five advice-oriented subreddits: r/Advice, r/CareerAdvice, r/FriendshipAdvice, r/AskMenAdvice, and r/AskWomenAdvice \cite{bhatia2025computational}. Each post is represented in a structured format that includes a textual description of the real-world \texttt{dilemma}, two alternative \texttt{options}, and the associated \texttt{costs} and \texttt{benefits} for each option. The dataset provides descriptions of everyday decision-making scenarios spanning personal, workplace, and social contexts, offering a rich and complex corpus for assessing LLMs’ value preferences.

Because understanding value trade-offs and LLM value preferences requires appropriate contextual granularity, we focus on four thematically coherent advice subreddits, namely r/CareerAdvice, r/FriendshipAdvice, r/AskMenAdvice, and r/AskWomenAdvice, instead of broad or general-purpose communities. The resulting dataset consists of 5,728 posts in total. Table~\ref{tab:dataset_stats} summarizes the basic statistics of the dataset.

\begin{table}[t]
\centering
\caption{Basic statistics of the dataset.}
\label{tab:dataset_stats}
\begin{tabular}{lcccc}
\toprule
\textbf{Subreddit} 
& r/careeradvice 
& r/FriendshipAdvice 
& r/AskMenAdvice 
& r/AskWomenAdvice \\
\midrule
\textbf{\# of posts} 
& 1,760 
& 484 
& 389 
& 3,095 \\
\midrule
\textbf{Total} & \multicolumn{4}{c}{\textbf{5,728}} \\
\bottomrule
\end{tabular}
\end{table}
 
\subsection{Value Extraction}
Based on the assumption that each option in a dilemma is motivated by a dominant core value \cite{rawls1951outline,jiang2025investigating}, our first step is to identify the value motivating each option. Recent large language models, such as GPT-4o, have demonstrated strong performance and reliability in text understanding and semantic judgment tasks \cite{ding2023gpt,gilardi2023chatgpt}, making them suitable for inferring latent values from natural language descriptions~\cite{sorensen2024value}. Following \citet{huang2025values}, we prompt GPT-4o to extract the core value associated with each option. Specifically, we provide GPT-4o with the \texttt{dilemma}, two alternative \texttt{options}, and corresponding \texttt{costs} and \texttt{benefits} for each option. We then ask the model to summarize, in one to four words, the core value motivating each option, along with a brief justification. The complete prompt is provided in the Appendix~\ref{appendix:value_prompt}.

\begin{figure}[t]
    \centering
    \includegraphics[width=0.80\columnwidth]{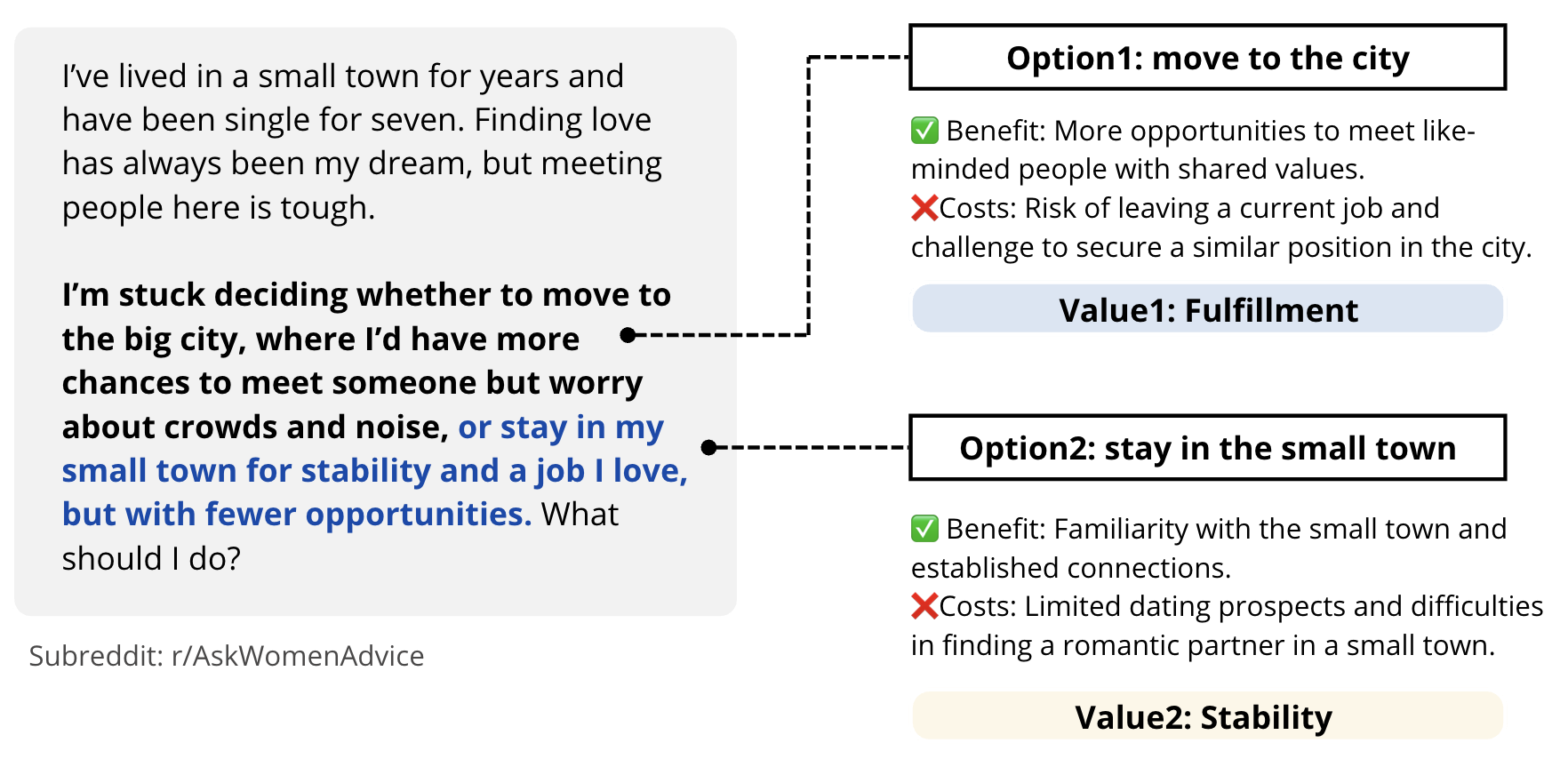}
    \caption{Example of a real-world dilemma sourced from r/AskWomenAdvice and the corresponding extracted values. Users describe a dilemma involving two mutually exclusive options (moving to a city vs.\ staying in a small town), each associated with different perceived benefits and costs. Our pipeline identifies the dominant core value motivating each option (\textit{Fulfillment} vs.\ \textit{Stability}), illustrating how everyday dilemmas encode value trade-offs. }
    \label{fig:examples_value_extract}
\end{figure}

To assess whether GPT-4o performs well in identifying values when provided with contextual information, we construct a validation set by randomly sampling 100 posts stratified from each subreddit, yielding a total of 400 posts. Since each post contains two options, each associated with an extracted value and justification, annotators evaluate 800 value assignments in total. Two authors independently review both the identified values and the accompanying reasoning, assessing whether they are appropriate given the contextual information. A value assignment is accepted as correct only when both annotators agree. Inter-annotator agreement is high (Cohen’s $\kappa = 0.833$), and 92\% of the 800 assignments pass validation, indicating that GPT-4o performs well in extracting values in our setting. We therefore apply GPT-4o to identify values for each option across the full dataset. Figure~\ref{fig:examples_value_extract} presents examples of the value extraction results. The resulting annotated dataset serves as the basis for all subsequent analyses in this study.

\subsection{LLM Decision Making}
Next, we prompt three large language models, GPT-4o, DeepSeek-V3.2-Exp, and Gemini-2.5-Flash, to make a choice when confronted with each dilemma. Specifically, we provide each model with the \texttt{dilemma} and the two alternative \texttt{options}, and ask it to choose one of the two. The model is required to return only the chosen option, without providing any additional explanation. We set the temperature to 0 to ensure stable outputs from the models. The complete prompt is provided in the Appendix~\ref{appendix:choice_prompt}.

To mitigate potential order effects, whereby a model may be biased toward selecting the first presented option, we conduct an additional experiment in which the order of the two options is reversed. The results show that approximately 92.5\% of model choices remain unchanged after the order is switched, indicating that model decisions are largely robust to the presentation order and are not driven by positional bias.

\subsection{Bottom-up Construction of a Hierarchical Value Framework}
After value extraction, we obtain 2,288 distinct values in total. To construct a hierarchical value framework in a bottom-up approach, we adopt the algorithm proposed in~\citet{huang2025values}, which proceeds as follows. It is governed by two hyperparameters: $n_{\text{top}}$, which specifies the desired number of clusters at the top level, and $L$, which denotes the total number of levels in the hierarchical framework. In addition, $n_{\text{base}}$ represents the number of initial clusters at the bottom level, which is 2,288 in our study.

To determine the number of clusters at each level, we follow the formula~\cite{huang2025values} that allocates the cluster count for each layer as a function of $n_{\text{top}}$, $L$, and $n_{\text{base}}$:
\[
\frac{n_l}{n_{l-1}} = \left( \frac{n_{\text{top}}}{n_{\text{base}}} \right)^{\frac{1}{L-1}} .
\]

For each level of the hierarchy, we follow the two-step refinement procedure introduced by \citet{huang2025values}:

\begin{itemize}
    \item We embed all clusters using \texttt{all-mpnet-base-v2} \cite{song2020mpnet} and apply \emph{k}-means clustering over these embeddings to group clusters into semantically coherent neighborhoods.
    \item For each resulting cluster, we prompt GPT-4o to generate a representative cluster name that corresponds to a higher-level value encompassing the subordinate values within the cluster. 
\end{itemize}
This process is repeated at each level until the desired number of top-level clusters, $n_{\text{top}}$, is reached.

Considering both the clarity of the framework presentation and the coverage of all values, we set $n_{\text{top}} = 4$ and $L = 4$ in our study. We also conduct a qualitative pass over all clusters and refine their names and descriptions in cases where they are insufficiently clear, overly verbose, or inaccurate. In addition, clusters with identical or highly similar labels are merged to avoid redundancy.
As a result, we obtain a hierarchical value framework consisting of four levels: 2{,}288 fine-grained values at Level$_0$ (bottom), 175 clustered values at Level$_1$, 33 intermediate values at Level$_2$, and 4 top-level value dimensions at Level$_3$ (top). At the top level, the framework comprises four core values: \textit{Exploration \& Growth}, \textit{Security \& Stability}, \textit{Achievement \& Impact}, and \textit{Benevolence \& Connection}. Further details can be found in Section~\ref{sec: framework}.

To assess \textbf{(i)} whether setting $n_{\text{top}} = 4$ sufficiently captures the overall landscape of the extracted values, and \textbf{(ii)} whether the clustering process is reasonable in the sense that lower-level values are accurately assigned to their corresponding higher-level clusters, we conduct an additional manual validation. Specifically, we randomly sample 20\% of the dataset ($N = 458$) and two authors independently assign each bottom-level value to one of the four top-level values, framing the task as a four-class classification problem. This design directly assesses the end-to-end mapping from the bottom level to the top level of the framework. Since the top-level values are themselves derived through an iterative bottom-up aggregation process, consistency at this level provides evidence that the hierarchical construction is reasonable. The two annotators agree on 423 out of 458 instances (92.4\% of the cases), with a Cohen’s $\kappa$ of 0.896, indicating high inter-annotator reliability. Among the instances on which both annotators agree, the assignments produced by the hierarchical clustering match the human judgments in 96.7\% of the cases, indicating that the proposed value hierarchy is reasonable and sufficiently covers the semantic space of the corpus.

\subsection{Quantifying Value Preferences of LLMs}
To quantify LLMs’ value preferences, we introduce a metric termed the \emph{winning rate}. 
The winning rate captures how often a given value is favored by a model when it is directly traded off against other values within the same dilemma.

For the analysis of RQ2 (value preferences of LLMs), we only consider the four values at the top level
(\textit{Exploration \& Growth}, \textit{Security \& Stability}, \textit{Achievement \& Impact}, and \textit{Benevolence \& Connection}), as they provide a compact and theoretically grounded
representation of the broader values while reducing sparsity and noise at finer-grained levels. We focus on pairwise comparisons between distinct values: for any two different values $i$ and $j$ ($i \neq j$), we define the \emph{pairwise winning rate} $R_{ij}$ as the proportion of times the model prefers value $i$ over value $j$, conditional on both values co-occurring in the same dilemma:

\begin{equation*}
R_{ij} =
\frac{
\text{\# of times the model prefers value } i
}{
\text{\# of dilemmas in which values } i \text{ and } j \text{ co-occur}
}
\end{equation*}

This formula normalizes the preference of value $i$ over value $j$ by the number of relevant trade-off situations, rather than by the total dataset size. To obtain an overall preference score for a value $i$, we aggregate its pairwise winning rates against the remaining three values.
Let $\{j, k, l\}$ denote the other three values distinct from $i$. The \emph{global winning rate} of value $i$, denoted as $R_i$, is defined as:

\begin{equation*}
R_i = \frac{1}{3} \left( R_{ij} + R_{ik} + R_{il} \right)
\end{equation*}

By construction, $R_i \in [0,1]$.
A value of $R_i > 0.5$ indicates that value $i$ is, on average, preferred over alternative values in direct trade-off situations, whereas $R_i < 0.5$ suggests the opposite.

For each LLM, we compute the winning rates of values at two different scales: an overall level and a subreddit level. First, to assess an LLM’s overall value preference, we aggregate dilemmas from all subreddits and compute winning rates on the combined dataset. This provides a global estimate of the model’s value tendencies across the entire corpus.
Second, to characterize an LLM’s value preferences within a specific community, we compute winning rates separately for each subreddit by considering all dilemmas originating from that subreddit. This allows us to capture how value trade-offs are resolved by the model within distinct communities.

\section{Results}
\subsection{The Hierarchical Value Framework}
\label{sec: framework}

\begin{figure}[t!]
    \centering
    \includegraphics[width=\linewidth]{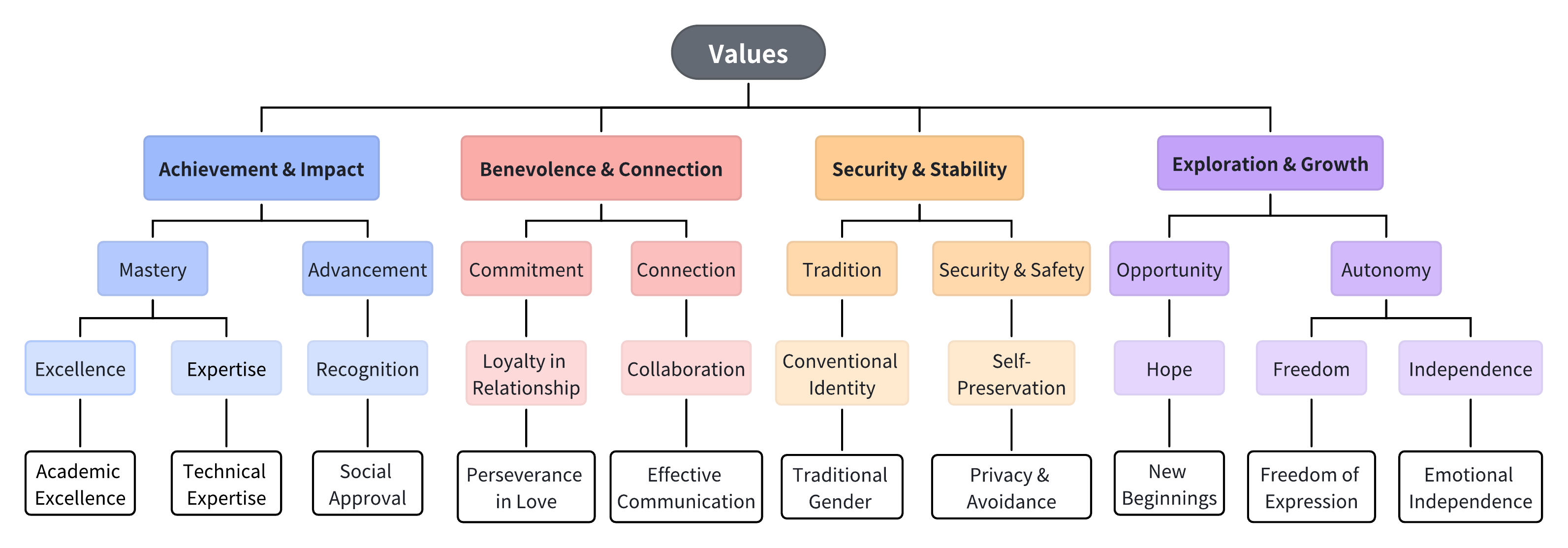}
    \caption{The hierarchical value framework constructed in this study through a bottom-up approach. The framework consists of four levels, with the top level comprising four higher-order value dimensions: \textit{Achievement \& Impact}, \textit{Benevolence \& Connection}, \textit{Security \& Stability}, and \textit{Exploration \& Growth}. Starting from fine-grained values extracted from real-world dilemmas (Level$_0$), semantically similar values are clustered into increasingly abstract value concepts at intermediate layers (Level$_1$ and Level$_2$), ultimately converging into four top-level value dimensions (Level$_3$). }
    \label{fig:values}
\end{figure}

Figure~\ref{fig:values} presents the hierarchical value framework constructed using our bottom-up approach. To enhance interpretability, the figure includes a set of representative examples for each cluster, rather than an exhaustive list of all values. These examples are intended to provide readers with an intuitive understanding of the values encompassed by each cluster. Within this framework, the bottom level consists of fine-grained values directly extracted from real-world dilemmas. Through three rounds of iterative \emph{k}-means clustering, similar values are progressively grouped into higher-level clusters, each representing a more abstract and general value concept. These clusters are further consolidated into four top-level values: \textit{Achievement \& Impact}, \textit{Benevolence \& Connection}, \textit{Security \& Stability}, and \textit{Exploration \& Growth}.

To illustrate this bottom-up approach, consider the value \textit{Perseverance in Love}. At the lowest level, it is identified as a concrete value motivating choices in real-world dilemmas. It is first grouped into the higher-level cluster \textit{Loyalty in Relationship}, which captures persistence and commitment within intimate bonds.
This cluster is then aggregated into the mid-level category \textit{Commitment}, reflecting sustained dedication in interpersonal relationships. Finally, \textit{Commitment} is subsumed under the higher-level value \textit{Benevolence \& Connection}, which captures broader concerns with social bonds, loyalty, and relational care.

At the top level of the framework, four core values are identified, which together capture the major landscape of value considerations present in real-world dilemmas. Specifically, each top-level value represents the following:

\begin{itemize}
    \item \textbf{Achievement \& Impact} reflects motivations related to competence, mastery, and achievement, emphasizing personal effectiveness, goal attainment, and the pursuit of meaningful impact.
    \item \textbf{Benevolence \& Connection} captures interpersonal and prosocial orientations, highlighting care for others, relational commitment, and the maintenance of meaningful social bonds.
    \item \textbf{Security \& Stability} represents preferences for safety and predictability, emphasizing the protection of personal well-being and social norms.
    \item \textbf{Exploration \& Growth} emphasizes openness to change and self-direction, reflecting motivations toward autonomy, novelty, and personal development.
\end{itemize}

It is noteworthy that the four values identified through our bottom-up approach exhibit substantial conceptual alignment with Schwartz’s theory of basic human values \cite{schwartz1992universals}. Schwartz’s framework originally identifies ten basic values, which are further organized along two higher-order dimensions: \textit{Self-Enhancement} versus \textit{Self-Transcendence}, and \textit{Openness to Change} versus \textit{Conservation}.
Within this theoretical structure, \textit{Achievement \& Impact} in our framework closely corresponds to \textit{Self-Enhancement}, while \textit{Benevolence \& Connection} aligns with \textit{Self-Transcendence}. Similarly, \textit{Security \& Stability} aligns with \textit{Conservation}, while \textit{Exploration \& Growth} aligns with \textit{Openness to Change}. This correspondence enhances the theoretical coherence between our empirically derived framework and well-established theories of human values, suggesting that value structures emerging from real-world dilemmas are consistent with broader patterns of human motivation. Overall, this hierarchical value framework serves as the foundation for our subsequent analyses of value trade-offs across different subreddits, as well as the value preferences exhibited by large language models.

\subsection{RQ1: Value Trade-offs in Advice-Oriented Subreddits}
\begin{figure}[t]
\centering

\begin{minipage}{0.48\columnwidth}
    \centering
    \includegraphics[width=0.8\linewidth]{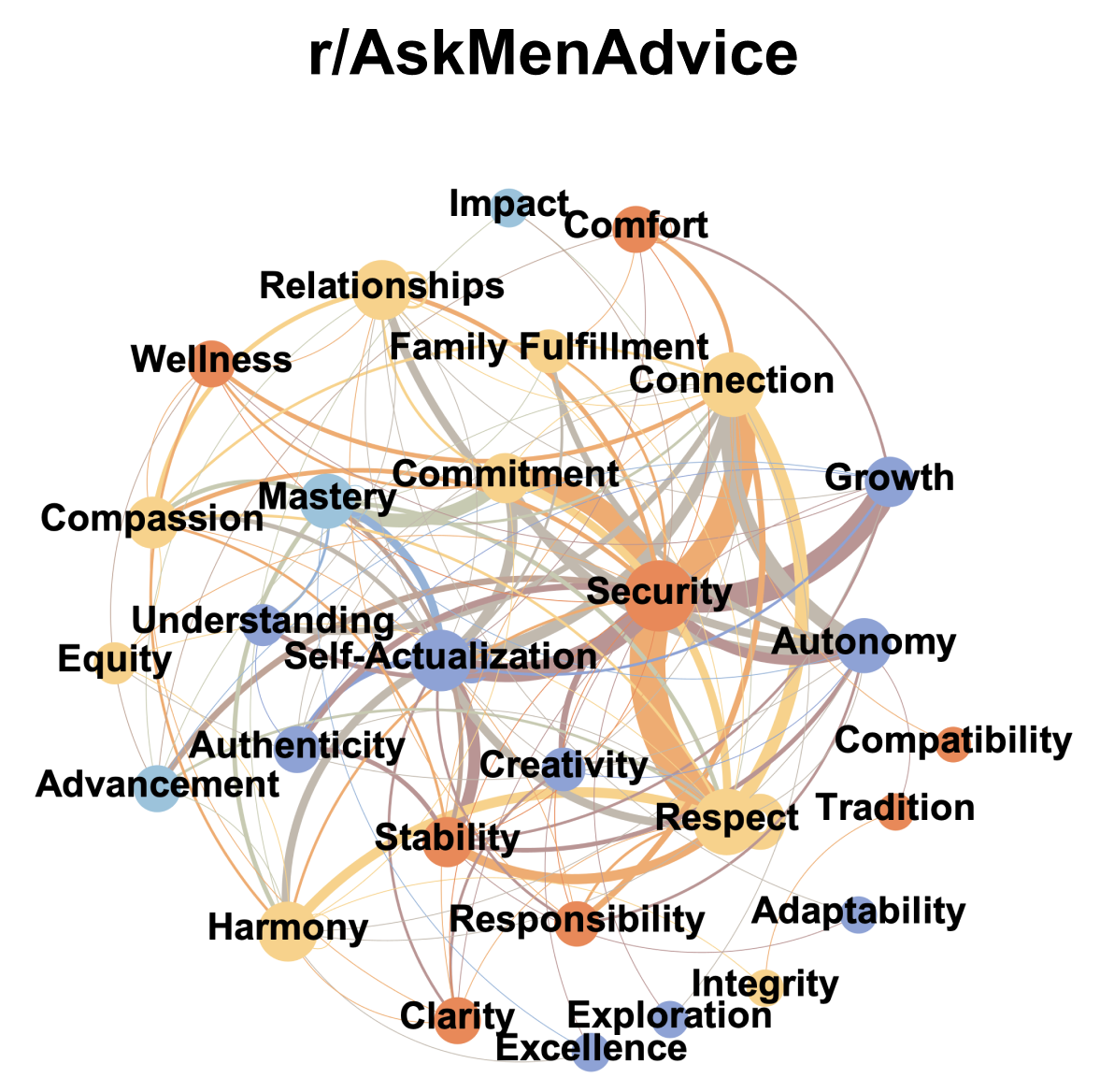}
    \vspace{2pt}
\end{minipage}
\hfill
\begin{minipage}{0.48\columnwidth}
    \centering
    \includegraphics[width=0.8\linewidth]{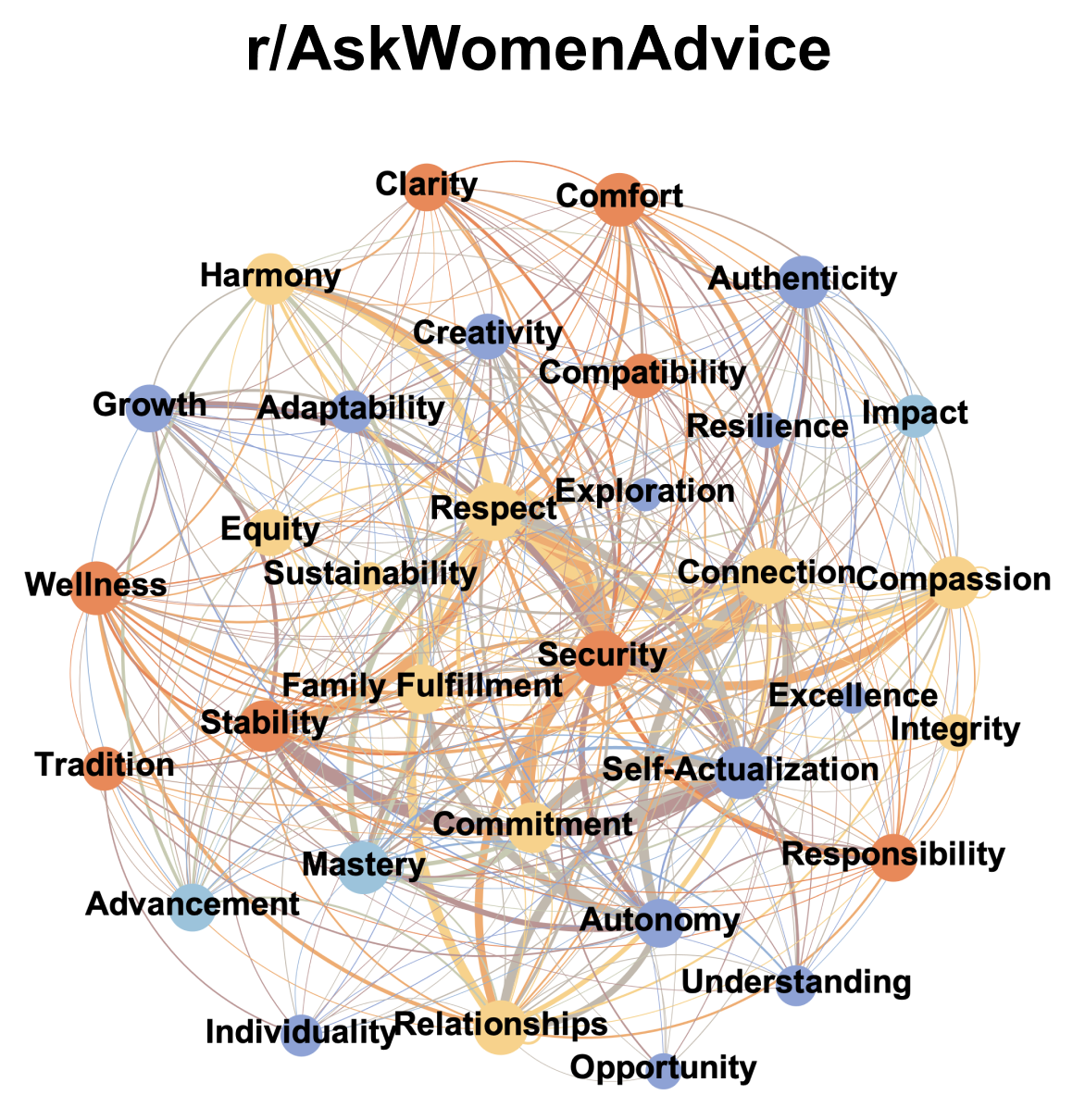}
    \vspace{2pt}
\end{minipage}

\vspace{5pt}

\begin{minipage}{0.48\columnwidth}
    \centering
    \includegraphics[width=0.8\linewidth]{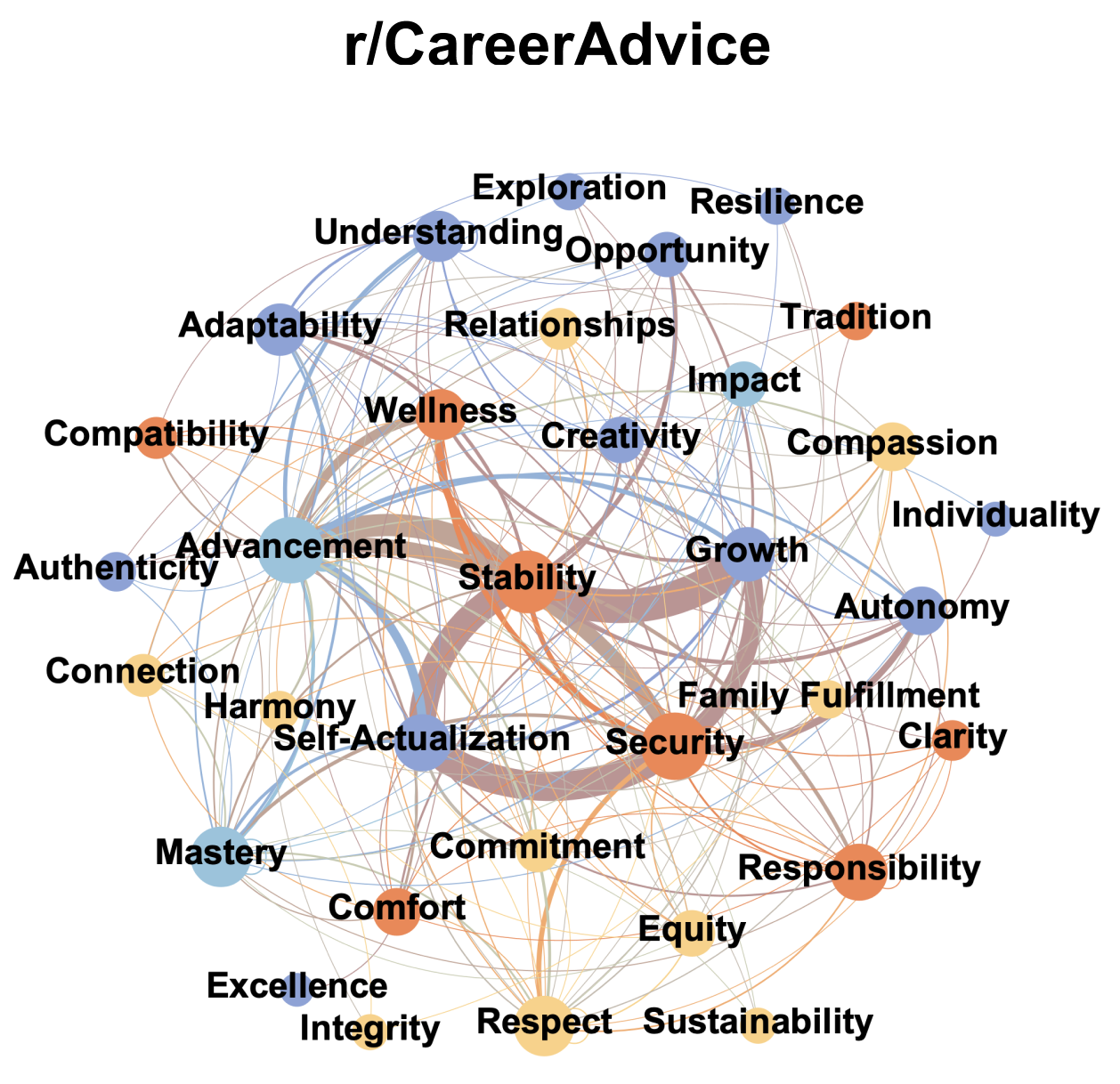}
    \vspace{2pt}
\end{minipage}
\hfill
\begin{minipage}{0.48\columnwidth}
    \centering
    \includegraphics[width=0.8\linewidth]{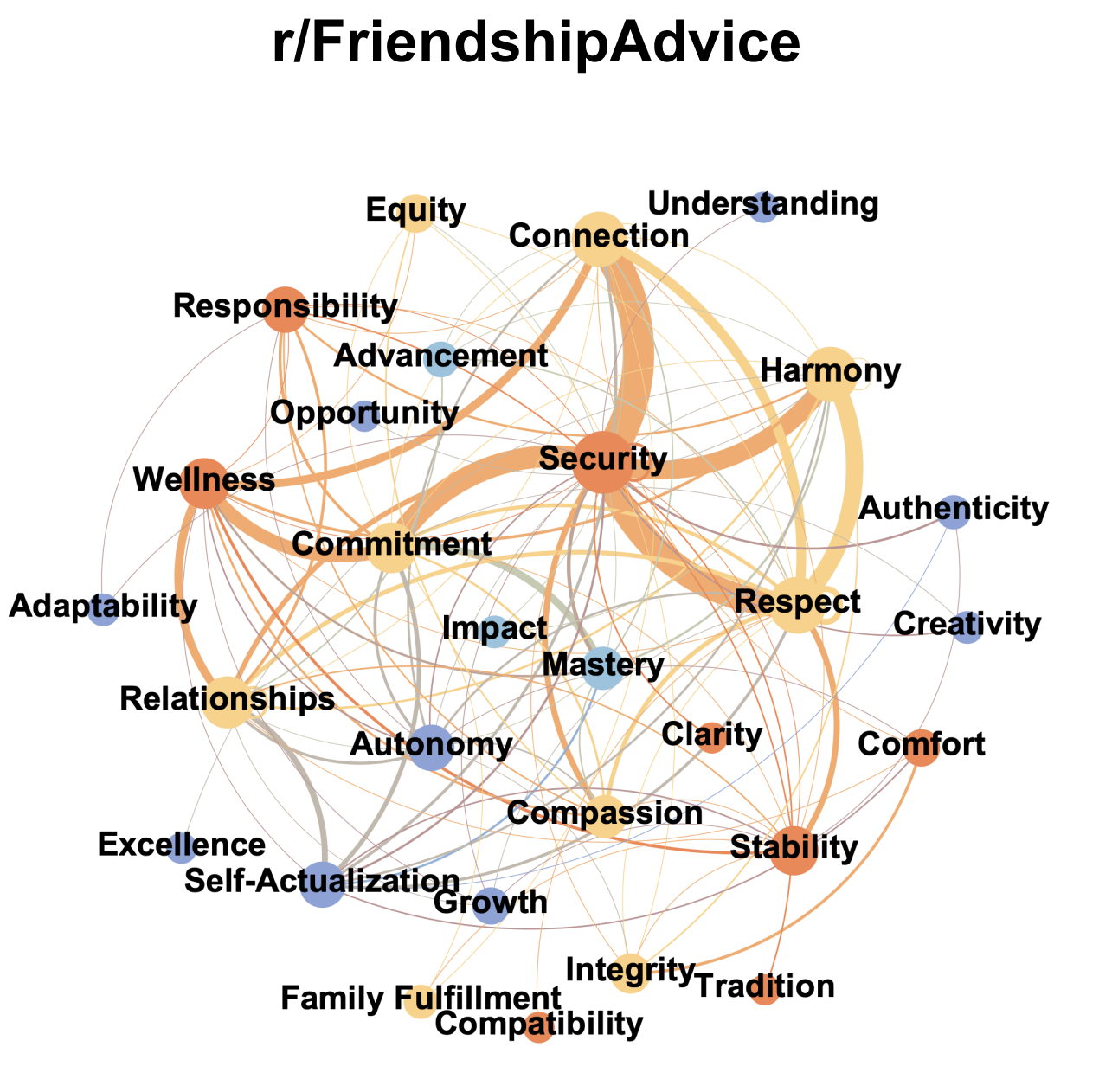}
    \vspace{2pt}
\end{minipage}

\caption{Value co-occurrence networks at Level$_2$ across four advice-oriented subreddits: r/AskMenAdvice, r/AskWomenAdvice, r/CareerAdvice, and r/FriendshipAdvice. Nodes represent values and edges indicate co-occurrence frequency within the same dilemma, with thicker edges reflecting stronger co-occurrence. Node colors correspond to the top-level values derived from our hierarchical framework: \textit{Security \& Stability} (orange), \textit{Benevolence \& Connection} (yellow), \textit{Exploration \& Growth} (blue), and \textit{Achievement \& Impact} (light blue). }
\label{fig:value_networks_2}
\end{figure}

To characterize value trade-offs that people frequently encounter in everyday dilemmas, we construct a value co-occurrence network for each subreddit. In each network, nodes represent values, while edges indicate the co-occurrence of two values within the same dilemma. The weight of an edge corresponds to the frequency with which the two values co-occur. Due to space constraints, we present value co-occurrence networks based on Level$_2$ values, as shown in Figure~\ref{fig:value_networks_2}. We focus on Level$_2$ because it contains a manageable number of values (33) that allow interpretable visualization while still capturing fine-grained value trade-offs. Networks at Level$_1$ (175 values) and Level$_3$ (4 values) are provided in Appendix~\ref{sec: value_networks}.

First, we observe that advice-seeking in the women-focused subreddit exhibits a significantly higher density than in the other subreddits. To determine whether this difference is driven by sample size or by the underlying structural properties of each subreddit, we apply bootstrap sampling ($B = 1{,}000$) to assess statistical significance. The results show that the network density of r/AskWomenAdvice is significantly higher than that of the other three subreddits ($p < 0.05$), indicating that \textit{\textbf{advice-seeking in women-focused communities involves a broader and more diverse set of value trade-offs}}.

\begin{figure}[t]
    \centering
    \includegraphics[width=0.8\columnwidth]{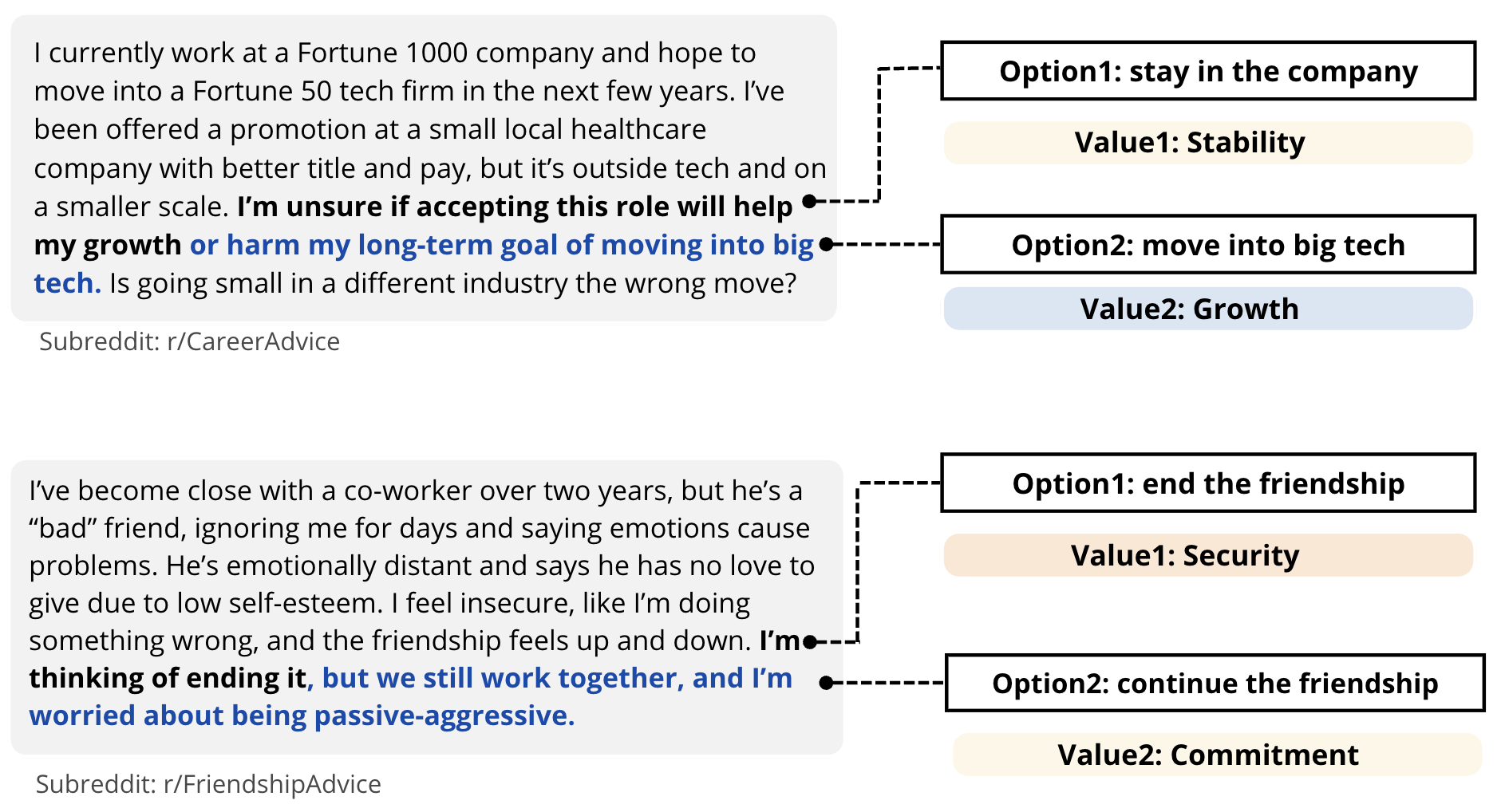}
    \caption{Illustrative examples of value trade-offs from r/CareerAdvice and r/FriendshipAdvice. In both dilemmas, users face two mutually exclusive options that involve competing values (e.g., \textit{Stability} vs.\ \textit{Growth} in r/CareerAdvice, \textit{Security} vs.\ \textit{Commitment} in r/FriendshipAdvice).}
    \label{fig:examples_value_tradeoff}
\end{figure}

Next, we rank value trade-offs by their edge weights and select the top five value trade-offs for each subreddit, as shown in Table~\ref{tab:top_tradeoffs}. Two key patterns emerge: \textbf{(i)} advice-seeking in men-, women-, and friendship-oriented subreddits exhibits highly similar value trade-off structures. These subreddits consistently emphasize tensions involving security, particularly \textit{Security vs.\ Respect}, \textit{Security vs.\ Connection}, and \textit{Security vs.\ Commitment}. In contrast, career-related advice forms a more distinct and weakly overlapping structure, with trade-offs primarily centered on values related to growth, self-actualization, and advancement; \textbf{(ii)} across contexts, frequently invoked values include security, respect, and commitment, which are often in tension with values emphasizing change, autonomy, or opportunity. This tension is especially pronounced in career-focused advice, where concerns about growth and advancement more frequently compete with stability-oriented values. Figure~\ref{fig:examples_value_tradeoff} presents examples of frequent value trade-offs in r/FriendshipAdvice and r/CareerAdvice.

\begin{table}[t]
\centering
\caption{Top five value trade-offs across the four advice-oriented subreddits, ranked by edge weight. Blue indicates trade-offs that appear in at least two subreddits.}
\label{tab:top_tradeoffs}
\resizebox{\columnwidth}{!}{%
\begin{tabular}{lcccc}
\toprule
\textbf{Rank} 
& \textbf{AskMenAdvice} 
& \textbf{AskWomenAdvice} 
& \textbf{CareerAdvice} 
& \textbf{FriendshipAdvice} \\
\midrule
1 &
\textcolor{blue}{Security vs. Respect} &
\textcolor{blue}{Security vs. Respect} &
Stability vs. Growth &
\textcolor{blue}{Security vs. Connection} \\
2 &
\textcolor{blue}{Security vs. Connection} &
\textcolor{blue}{Security vs. Connection} &
\textcolor{blue}{Security vs. Self-Actualization} &
\textcolor{blue}{Security vs. Respect} \\
3 &
\textcolor{blue}{Security vs. Growth} &
\textcolor{blue}{Security vs. Commitment} &
Stability vs. Self-Actualization &
\textcolor{blue}{Security vs. Commitment} \\
4 &
\textcolor{blue}{Security vs. Commitment} &
Security vs. Stability &
Stability vs. Advancement &
Harmony vs. Security \\
5 &
\textcolor{blue}{Security vs. Self-Actualization} &
Connection vs. Respect &
\textcolor{blue}{Security vs. Growth} &
Harmony vs. Respect \\
\bottomrule
\end{tabular}%
}
\end{table}

\paragraph{Summary.}
We present three key findings regarding the value trade-offs in four advice-oriented subreddits. First, a greater diversity of value trade-offs emerges in r/AskWomenAdvice compared to the other subreddits. Second, advice-seeking in men-, women-, and friendship-oriented subreddits reveals largely overlapping value trade-off patterns. Third, security emerges as a consistently salient value across subreddits and frequently stands in tension with values oriented toward change, autonomy, or opportunity. This tension is particularly evident in a career-focused subreddit.

\subsection{RQ2: Value Preferences of LLMs}
Building on the metric introduced to quantify the value preferences of large language models, our analysis has two objectives. First, we examine the overall value preferences of LLMs and identify common patterns across different models. Second, we take a closer look at how value preferences vary across subreddits, providing a more fine-grained analysis of contextual differences in model value preferences.

To assess the significance of the observed differences in the value preferences of LLMs, we conduct non-parametric bootstrap tests. First, when assessing overall value preferences (aggregated across all subreddits) and within-subreddit value preferences, the goal is to determine whether an LLM prefers one value over another within the same context. For each bootstrap iteration $b \in \{1,\dots, B\}$, we compute the winning-rate statistic $G_b(v)$ for each value $v$, and then apply the bootstrap test using the empirical sampling distribution
\[
G_b(v_1) - G_b(v_2).
\]

Second, when assessing between-subreddit value preferences, the goal is different: we test whether the same value is preferred to the same degree across two advice contexts (subreddits). Let $G_b^{\text{sub}}(v)$ denote the winning rate of value $v$ restricted to dilemmas from subreddit $\text{sub}$. We apply the bootstrap test using the empirical sampling distribution
\[
G_b^{\text{sub}1}(v) - G_b^{\text{sub}2}(v).
\]

Two-sided $p$-values are computed using the empirical tail probability
\[
p = 2 \cdot \min\big(\mathbb{P}(\Delta \le 0), \mathbb{P}(\Delta \ge 0)\big),
\]
where $\Delta$ denotes the bootstrap difference. This procedure makes no distributional assumptions and is robust to non-normality of the winning-rate statistic.

\subsubsection{Overall Value Preferences of LLMs}
\begin{figure*}[t!]
    \centering
    \includegraphics[width=0.82\linewidth]{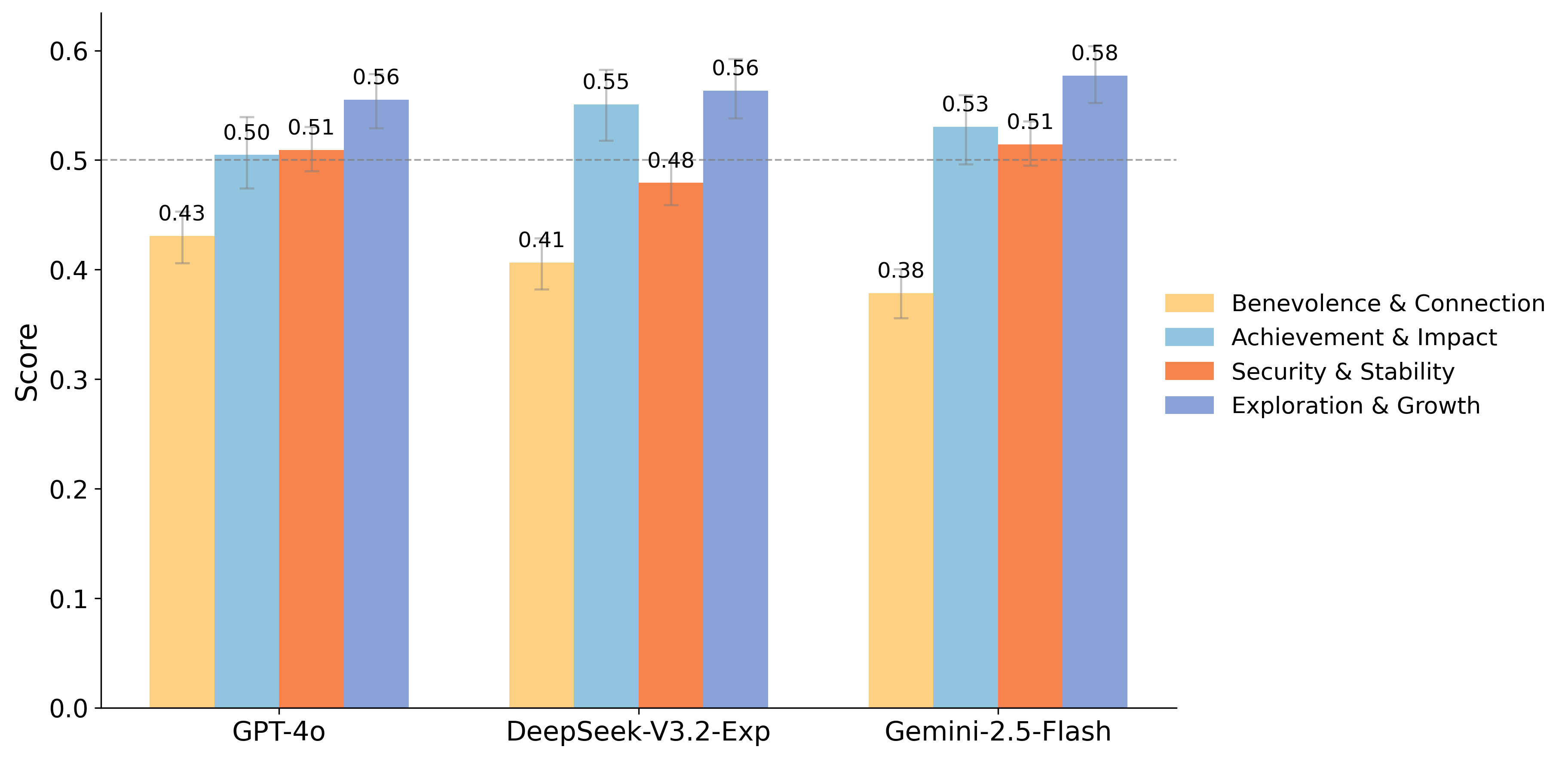}
    \caption{Value preferences across three large language models (GPT-4o, DeepSeek-V3.2-Exp, and Gemini-2.5-Flash). Scores reflect the winning rates of each value. Error bars indicate 95\% bootstrap confidence intervals ($B=1{,}000$). Across models, \textit{Benevolence \& Connection} receives the lowest winning rate, while \textit{Exploration \& Growth} consistently receives the highest winning rate.}
    \label{fig:total-llm}
\end{figure*}

Figure \ref{fig:total-llm} presents the overall value preferences of GPT-4o, DeepSeek-V3.2-Exp, and Gemini-2.5-Flash. Three consistent patterns emerge across all three models: \textbf{(i)} \textit{\textbf{\textit{Benevolence \& Connection} is the least preferred value}}, exhibiting a significantly lower winning rate than the other three values (all models: $p < 0.001$); \textbf{(ii)}\ \textit{\textbf{\textit{Exploration \& Growth} is consistently more preferred than both \textit{Security \& Stability}}} (DeepSeek-V3.2-Exp and Gemini-2.5-Flash: $p < 0.001$; GPT-4o: $p = 0.018$) \textbf{\textit{and Benevolence \& Connection}} (all models: $p < 0.001$); \textbf{(iii)} no statistically significant differences are observed between \textit{Exploration \& Growth} and \textit{Achievement \& Impact} (GPT-4o: $p = 0.05$; DeepSeek-V3.2-Exp: $p = 0.652$; Gemini-2.5-Flash: $p = 0.066$). In addition, for DeepSeek-V3.2-Exp, \textit{Achievement \& Impact} exhibits a significantly higher winning rate than \textit{Security \& Stability} ($p = 0.002$). Full results of the significance tests can be found in Appendix~\ref{sec: Overall Value}.

\subsubsection{Within-Subreddit Value Preferences of LLMs}
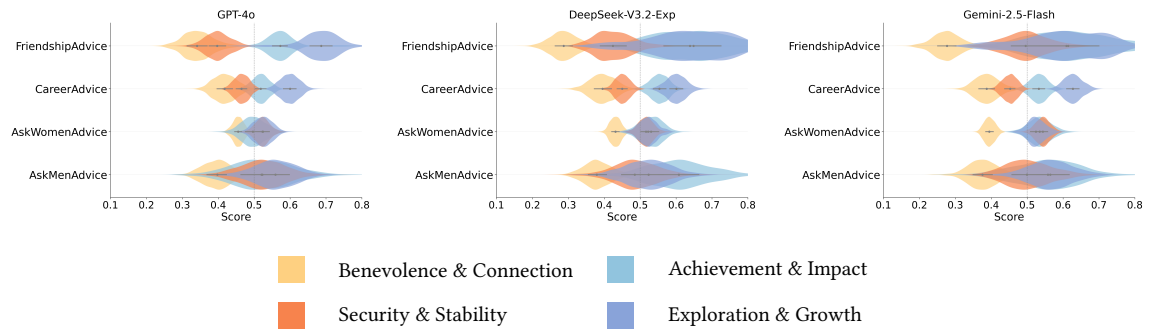
\begin{figure*}[t]
\centering

\begin{minipage}{0.32\textwidth}
    \centering
    \includegraphics[width=\linewidth]{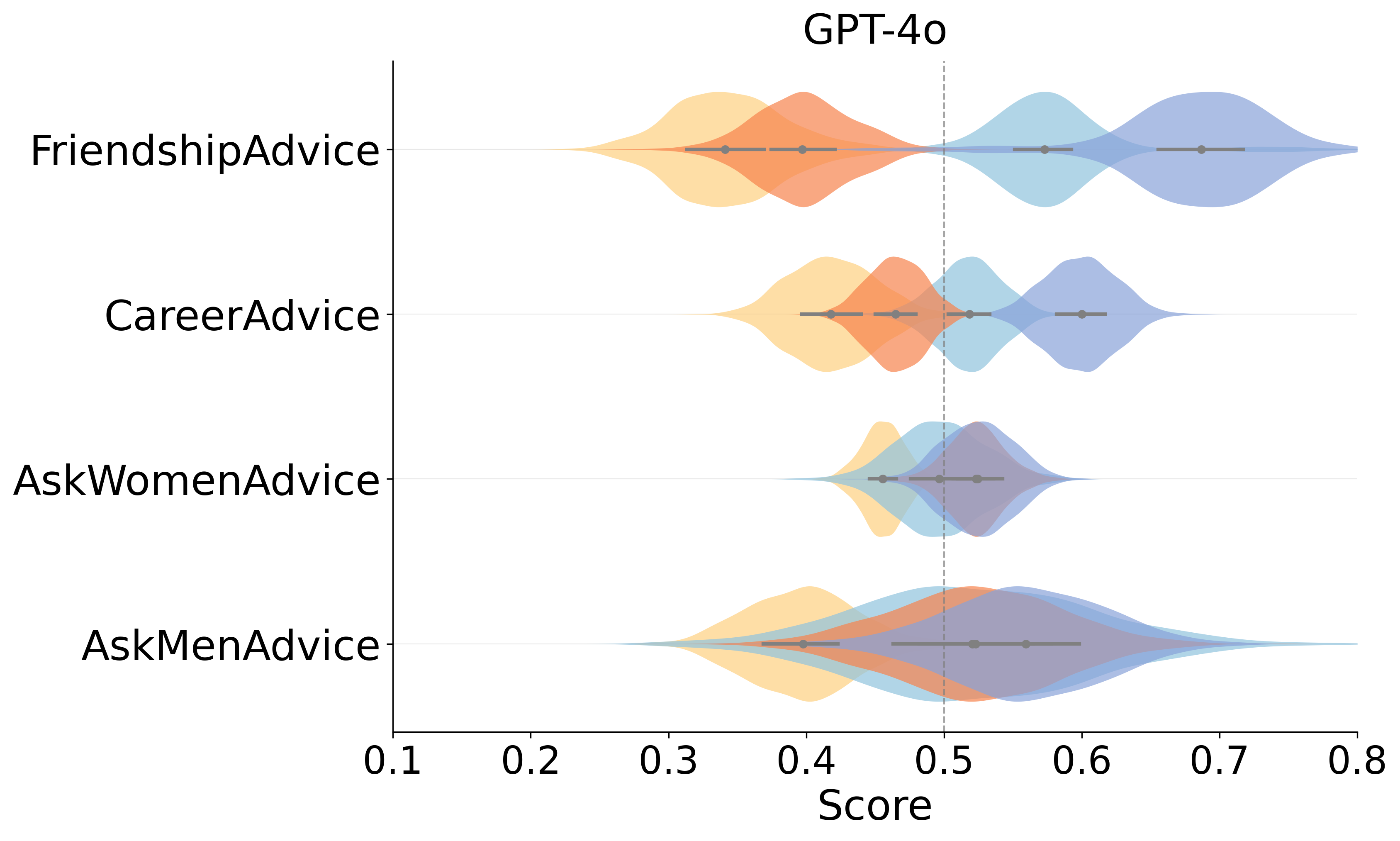}
\end{minipage}
\hfill
\begin{minipage}{0.32\textwidth}
    \centering
    \includegraphics[width=\linewidth]{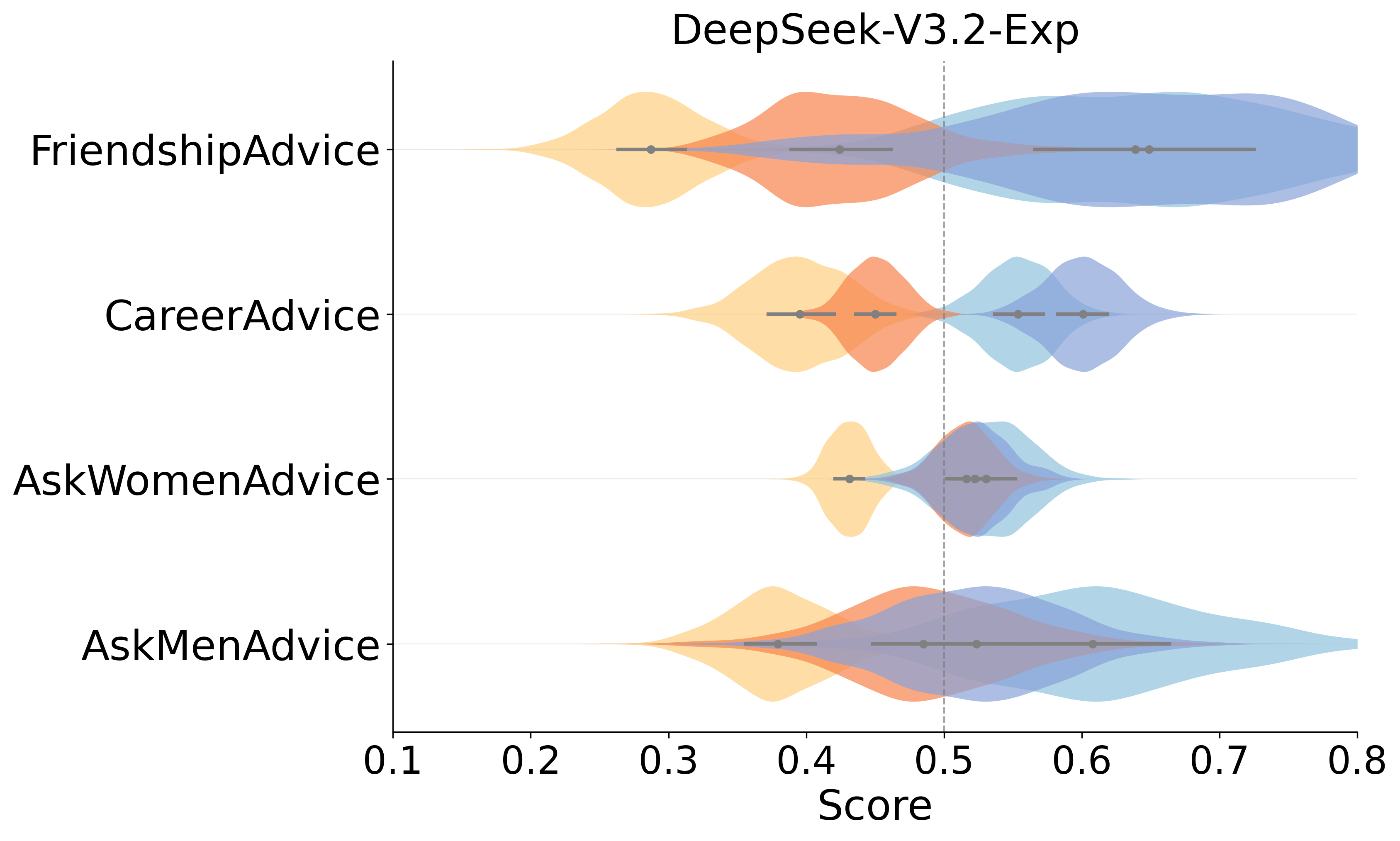}
\end{minipage}
\hfill
\begin{minipage}{0.32\textwidth}
    \centering
    \includegraphics[width=\linewidth]{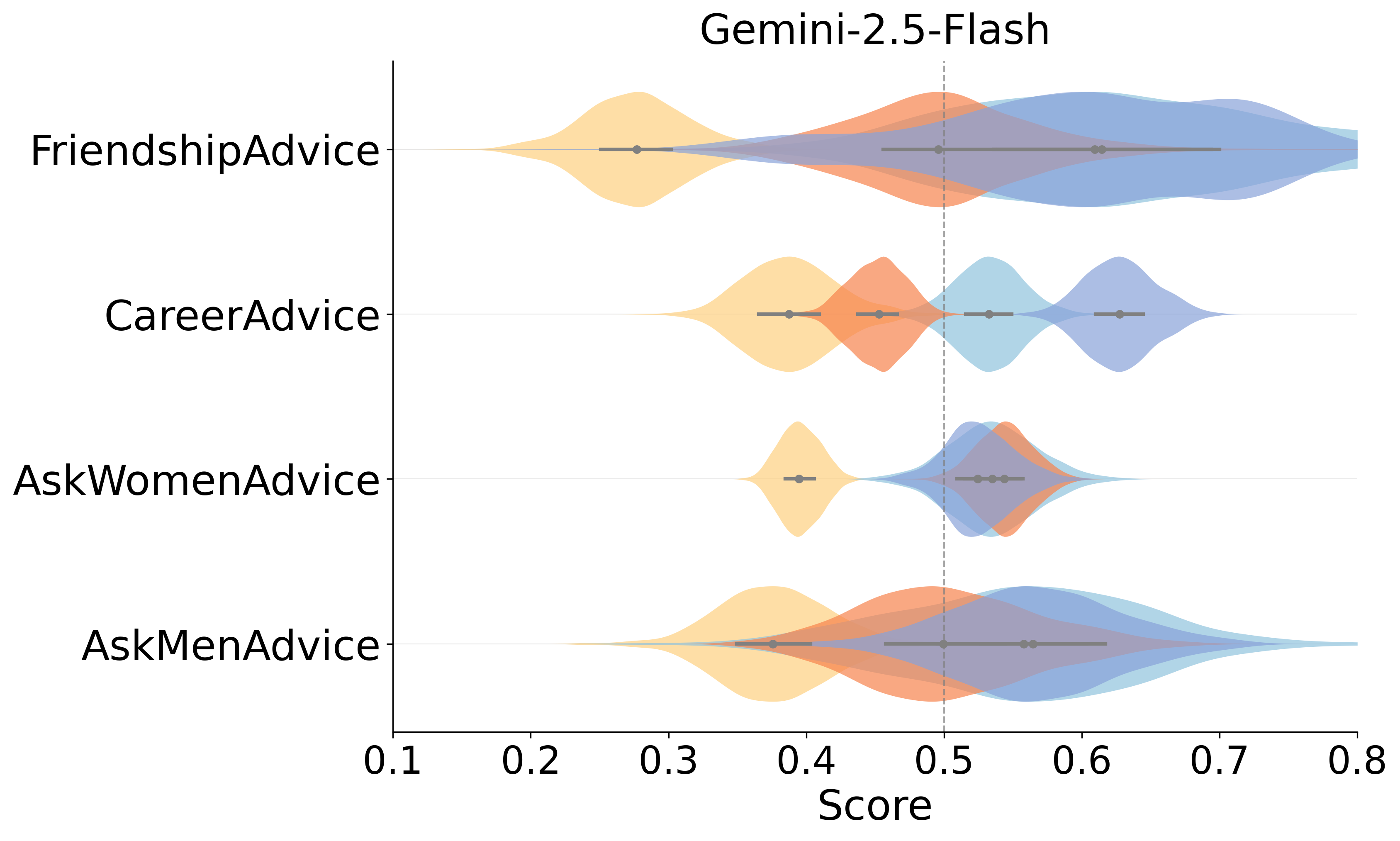}
\end{minipage}

\vspace{6pt}

\begin{tikzpicture}[font=\small]
    \definecolor{Benevolence}{HTML}{FED081}
    \definecolor{Achievement}{HTML}{91C4DE}
    \definecolor{Security}{HTML}{F7834D}
    \definecolor{Exploration}{HTML}{89A3D9}

    \matrix[row sep=4pt, column sep=10pt]{
        \node[fill=Benevolence,  minimum width=0.35cm, minimum height=0.35cm] {}; &
        \node[anchor=west]{Benevolence \& Connection}; &
        \node[fill=Achievement, minimum width=0.35cm, minimum height=0.35cm] {}; &
        \node[anchor=west]{Achievement \& Impact}; \\
        \node[fill=Security,     minimum width=0.35cm, minimum height=0.35cm] {}; &
        \node[anchor=west]{Security \& Stability}; &
        \node[fill=Exploration,  minimum width=0.35cm, minimum height=0.35cm] {}; &
        \node[anchor=west]{Exploration \& Growth}; \\
    };
\end{tikzpicture}

\vspace{2pt}

\caption{Violin plots depicting value preferences of three large language models across four advice-oriented subreddits. Each subplot visualizes the bootstrap-sampled winning rates ($B=1{,}000$) of four core values, indicating how often a value “wins” against competing values in real-world dilemmas. Central dots denote medians, horizontal bars indicate interquartile ranges (IQR), and the vertical dashed line marks the neutrality baseline ($0.5$), where a value wins as often as it loses.}

\label{fig:llm_values}
\end{figure*}

Following the same approach, we conduct a more fine-grained analysis of value preferences for each large language model across different subreddits, as shown in Figure~\ref{fig:llm_values}.

For each model, we first examine value preferences within each subreddit. The bootstrapped significance test reveals that all three models \textbf{\textit{consistently prioritize \textit{Exploration \& Growth} over \textit{Benevolence \& Connection} across all subreddits}}. Specifically, for GPT-4o this preference is significant in r/AskMenAdvice ($p=0.036$), r/AskWomenAdvice ($p=0.018$), and in both r/CareerAdvice and r/FriendshipAdvice ($p<0.001$). For DeepSeek-V3.2-Exp, it is significant in r/AskWomenAdvice ($p=0.002$), r/CareerAdvice ($p<0.001$), and r/FriendshipAdvice ($p=0.01$), but not in r/AskMenAdvice ($p=0.10$). For Gemini-2.5-Flash, significance holds in r/AskMenAdvice and r/FriendshipAdvice ($p=0.01$), and in r/AskWomenAdvice and r/CareerAdvice ($p<0.001$). Another noteworthy pattern is that DeepSeek-V3.2-Exp also prioritizes \textit{Achievement \& Impact} over \textit{Benevolence \& Connection} across all subreddits (r/AskMenAdvice: $p=0.028$; r/AskWomenAdvice: $p=0.018$; r/CareerAdvice: $p=0.002$; r/FriendshipAdvice: $p=0.004$). A similar pattern holds for Gemini-2.5-Flash in three subreddits (r/AskWomenAdvice: $p<0.001$; r/CareerAdvice and r/FriendshipAdvice: $p=0.002$), and for GPT-4o in two subreddits (r/CareerAdvice: $p=0.03$; r/FriendshipAdvice: $p=0.002$). Full results of the significance tests can be found in Appendix~\ref{sec: Within-Subreddit Value}. 

\subsubsection{Between-Subreddit Value Preferences of LLMs}
We then examine value preferences between different subreddits. Bootstrapped significance testing indicates that, for all three models, \textit{\textbf{\textit{Security \& Stability} has a significantly higher winning rate in r/AskWomenAdvice compared to r/CareerAdvice}} (GPT-4o: $p=0.044$; DeepSeek-V3.2-Exp: $p=0.026$; Gemini-2.5-Flash: $p=0.004$), whereas \textbf{\textit{\textit{Exploration \& Growth} shows a significantly lower winning rate in r/AskWomenAdvice than in r/CareerAdvice}} (GPT-4o: $p=0.038$; DeepSeek-V3.2-Exp: $p=0.03$; Gemini-2.5-Flash: $p=0.006$).  Full results of the significance tests can be found in Appendix~\ref{sec: Between-Subreddit Value}. 

\paragraph{Summary.} We present two main findings regarding the value preferences of LLMs. First, across all three models, \textit{Benevolence \& Connection} is consistently the least preferred value, and \textit{Exploration \& Growth} is systematically preferred over it. Notably, this pattern holds both at the overall level and at the within-subreddit level, indicating that core value orientations are robust to contextual variation. Second, despite this overall stability, context-specific shifts emerge: models place higher weight on \textit{Security \& Stability} in women-focused advice, whereas \textit{Exploration \& Growth} becomes more prominent in career-related contexts.

\section{Discussion and Conclusions}
In this study, we examine how large language models (LLMs) respond to everyday dilemmas and explore the values that motivate their choices. To do so, we leverage a curated dataset of 5,728 real-world dilemmas sourced from four advice-oriented subreddits and inductively construct a hierarchical value framework using a bottom-up approach. This framework serves two purposes. First, it allows us to uncover the value trade-off structures that underlie advice-seeking on social media platforms. Second, it enables us to investigate how LLMs navigate such trade-offs across different advice contexts.

\paragraph{The Hierarchical Value Framework} Our examination of LLM value preferences aligns with prior work that analyzes value preferences through dilemma-based evaluation (e.g., \cite{DBLP:conf/iclr/ChiuJ025,jiang2025investigating}). Similar to these studies, we rely on the assumption that values can be inferred from an agent’s choices when facing trade-offs~\cite{rawls1951outline}. However, prior work typically employs hand-crafted dilemmas and adopts a top-down value framework. While top-down approaches offer interpretability, they are often limited in capturing the diversity and granularity of value expressions observed in real-world settings. To address these limitations, we adopt a bottom-up approach grounded in everyday advice-seeking interactions. Notably, the top-level values of our constructed framework align with Schwartz’s higher-order value dimensions~\cite{schwartz1992universals}, demonstrating theoretical coherence. This bottom-up framework contributes to the literature in two ways. First, it provides an ecologically grounded representation of the value space that underlies everyday dilemmas. Second, the framework may facilitate both theoretical investigations (e.g., value inference and pluralism) and practical evaluations of LLMs’ value-related behavior.

\paragraph{Contextual Differences in Value Trade-offs} When examining the value trade-offs that underlie advice-seeking on social media platforms, our analysis highlights that women-focused advice involves a broader and more diverse set of value trade-offs. This is consistent with prior research suggesting that women often navigate multiple and intersecting role demands (e.g., work and family) that require balancing personal goals, relational commitments, and social expectations, thereby introducing a wider range of value trade-offs~\cite{carr2002psychological}. In addition, we find that women-, men-, and friendship-oriented subreddits share highly overlapping trade-off structures. A qualitative inspection shows that “relationship management” is the predominant theme across these subreddits, which suggests that advice communities focused on maintaining interpersonal relationships tend to emphasize social harmony and relationship maintenance, whereas career advice prioritizes growth and autonomy. This highlights the context-dependent nature of value negotiation in advice-seeking.

\paragraph{Preference for \textit{Exploration \& Growth} over \textit{Benevolence \& Connection}} Another noteworthy observation is that LLMs consistently prioritize \textit{Exploration \& Growth} over \textit{Benevolence \& Connection} across models and contexts. As more people not only seek advice from LLMs~\cite{cheong2024not, schneiders2025objection} but also come to rely on them as a substitute for interpersonal counsel~\cite{malfacini2025impacts, zhang2025rise}, these value preferences of LLMs may shape how individuals navigate trade-offs in their personal lives. This trend raises broader societal implications. If LLM-mediated advice increasingly favors exploration, autonomy, or opportunity over care, connection, or relational stability, it may contribute to subtle shifts in how normative decisions are made at scale. Our findings highlight the possibility that LLMs not only provide informational assistance, but also encode implicit value orientations that may influence downstream decision-making and interpersonal dynamics.

\paragraph{Limitations and Future Work}
We acknowledge several limitations in our research, which will guide future research. First, our study simplifies each option as being driven by a single dominant value, and we rely on GPT-4o to automatically infer that value. In practice, however, decision-making rarely reflects a one-to-one mapping between actions and values; rather, a single option may embody multiple competing or complementary value considerations. Future work could further relax this assumption and model multi-value compositions within a single option, paving the way for richer representations of value conflict and choice dynamics. Second, our evidence for value homogenization relies on revealed preferences inferred from forced-choice decisions within a limited set of models and prompting strategies. Future work should examine whether these patterns generalize across broader model families and prompting conditions. Beyond methodological extensions, future work should also examine downstream implications. If LLM-mediated advice systematically amplifies certain values while suppressing others, this may affect whose viewpoints feel represented, which recommendations users perceive as legitimate, and how normative judgments shift at scale. Studying user outcomes, perceived value alignment, and potential harms in real advice-seeking interactions would help translate descriptive findings about value preferences into actionable design and governance recommendations.

\section{Ethical Statement}
This study relies on publicly available data collected from Reddit, specifically from advice-oriented subreddits where users voluntarily share dilemmas and seek guidance from others. We use an existing dataset in which posts have been curated, structured, and anonymized by prior work. We do not attempt to identify individual users nor infer personal identities, and all analyses are conducted at the aggregate level. Nevertheless, advice-seeking posts often involve sensitive personal, relational, or professional situations. To mitigate potential risks, we focus on value trade-offs and aggregate patterns rather than individual cases, and refrain from making claims about specific users or communities beyond observed statistical trends. As a result, we consider the risk of direct harm to individuals to be low. 

We also acknowledge broader ethical risks associated with this line of work. In particular, identifying patterns of value homogenization raises concerns about whose values are amplified or marginalized in AI value preferences. These concerns underscore the need for future work that incorporates diverse human judgments, cross-cultural perspectives, and participatory evaluation when designing or deploying advice-giving AI systems.

\textbf{Generative AI Usage Statement}. During the preparation of this work, the authors used ChatGPT for text refinement and code review. The authors reviewed and edited the content as needed and take full responsibility for the content of the publication.

\bibliographystyle{ACM-Reference-Format}
\bibliography{sample-base}

\newpage

\appendix

\section{Prompt Design Details}
\subsection{Prompt for Value Extraction}
\label{appendix:value_prompt}
We extract values from real-world dilemmas using the prompt shown in Figure~\ref{fig:prompt_value_extract}.

\begin{figure*}[h]
    \centering
    \begin{tikzpicture}[font=\small]
        \node[
            draw,
            fill=gray!10,
            rounded corners,
            inner sep=10pt
        ] {
            \begin{minipage}{0.95\textwidth}
                \raggedright
                \begin{tabular}{l p{0.86\textwidth}}
                    \texttt{INPUT} &
                    Your task is to analyze an everyday dilemma and identify the single core value that underlies each of the two alternative options presented.\newline
                    You will be given and must analyze the situation, the option, and its associated benefits and costs.\newline
                    Do not ask any questions and only return what you are asked to output.\newline\newline
                    Situation: \{\textit{post\_content}\}\newline
                    --- Option 1 ---\newline
                    Choice: \{\textit{first\_option}\}\newline
                    Benefits:\newline
                    1. \{\textit{benefits1\_1}\}\newline
                    2. \{\textit{benefits1\_2}\}\newline
                    3. \{\textit{benefits1\_3}\}\newline
                    Costs:\newline
                    1. \{\textit{costs1\_1}\}\newline
                    2. \{\textit{costs1\_2}\}\newline
                    3. \{\textit{costs1\_3}\}\newline
                    --- Option 2 ---\newline
                    Choice: \{\textit{second\_option}\}\newline
                    Benefits:\newline
                    1. \{\textit{benefits2\_1}\}\newline
                    2. \{\textit{benefits2\_2}\}\newline
                    3. \{\textit{benefits2\_3}\}\newline
                    Costs:\newline
                    1. \{\textit{costs2\_1}\}\newline
                    2. \{\textit{costs2\_2}\}\newline
                    3. \{\textit{costs2\_3}\}\newline\newline
                    \textbf{INSTRUCTIONS:}\newline
                    For Option 1, identify the single most dominant core value that this choice seeks to uphold.\newline
                    For Option 2, independently identify the single most dominant core value that this choice seeks to uphold.\newline
                    Focus solely on the core value that motivates the choice.\newline
                    If any Option (1 or 2) lacks benefits/costs data, return ``NaN'' for its value.\newline
                    \textbf{OUTPUT REQUEST:}\newline
                    Write 2--4 sentences of reasoning explaining the values for both Option 1 and Option 2.\newline
                    Summarize the core values in 1--4 words each.\newline
                    If no values are clearly demonstrated, write ``None''.\newline
                    \textbf{OUTPUT FORMAT:}\newline
                    [2--4 sentences of reasoning]\newline
                    Option 1 Value: [1--4 words]\newline
                    Option 2 Value: [1--4 words]
                \end{tabular}
            \end{minipage}
        };
    \end{tikzpicture}
    \caption{Prompt format used for value extraction.}
    \label{fig:prompt_value_extract}
\end{figure*}

\clearpage
\subsection{Prompt for LLM Decision Making}
\label{appendix:choice_prompt}
We prompt large language models to make a choice between two options using the template shown in Figure~\ref{fig:prompt_choice}.

\begin{figure*}[h]
    \centering
    \begin{tikzpicture}[font=\small]
        \node[
            draw,
            fill=gray!10,
            rounded corners,
            inner sep=10pt
        ] {
            \begin{minipage}{0.95\textwidth}
                \raggedright
                \begin{tabular}{l p{0.86\textwidth}}
                    \texttt{INPUT} &
                    Your task is to analyze an everyday dilemma, review the two options presented, and then choose the single best option.\newline
                    The choice must be based on a comprehensive assessment of the situation and the two options.\newline
                    Do not ask any questions and only return what you are asked to output.\newline
                    Respond with only one word: ``Option~1'' or ``Option~2''.\newline

                    Situation: \{\textit{content}\}\newline
                    Option~1: \{\textit{first\_option}\}\newline
                    Option~2: \{\textit{second\_option}\}\newline
                \end{tabular}
            \end{minipage}
        };
    \end{tikzpicture}
    \caption{Prompt format used for LLM decision making.}
    \label{fig:prompt_choice}
\end{figure*}

\clearpage
\section{Additional Value Co-occurrence Networks}
\label{sec: value_networks}
\subsection{Value Co-occurrence Networks at Level$_3$}

\begin{figure}[h]
\centering

\begin{minipage}{0.48\columnwidth}
    \centering
    \includegraphics[width=0.85\linewidth]{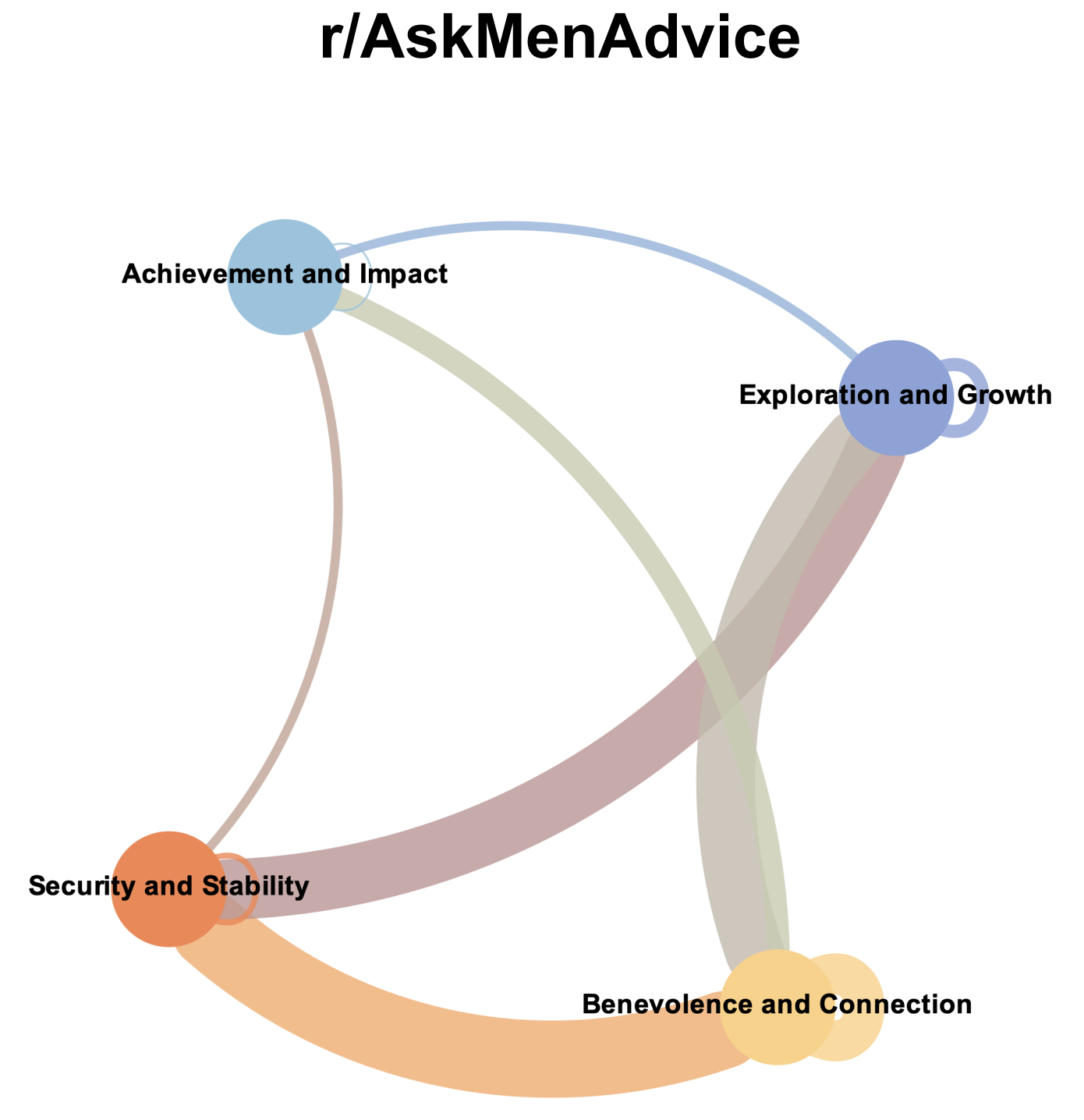}
    \vspace{2pt}
\end{minipage}
\hfill
\begin{minipage}{0.48\columnwidth}
    \centering
    \includegraphics[width=0.85\linewidth]{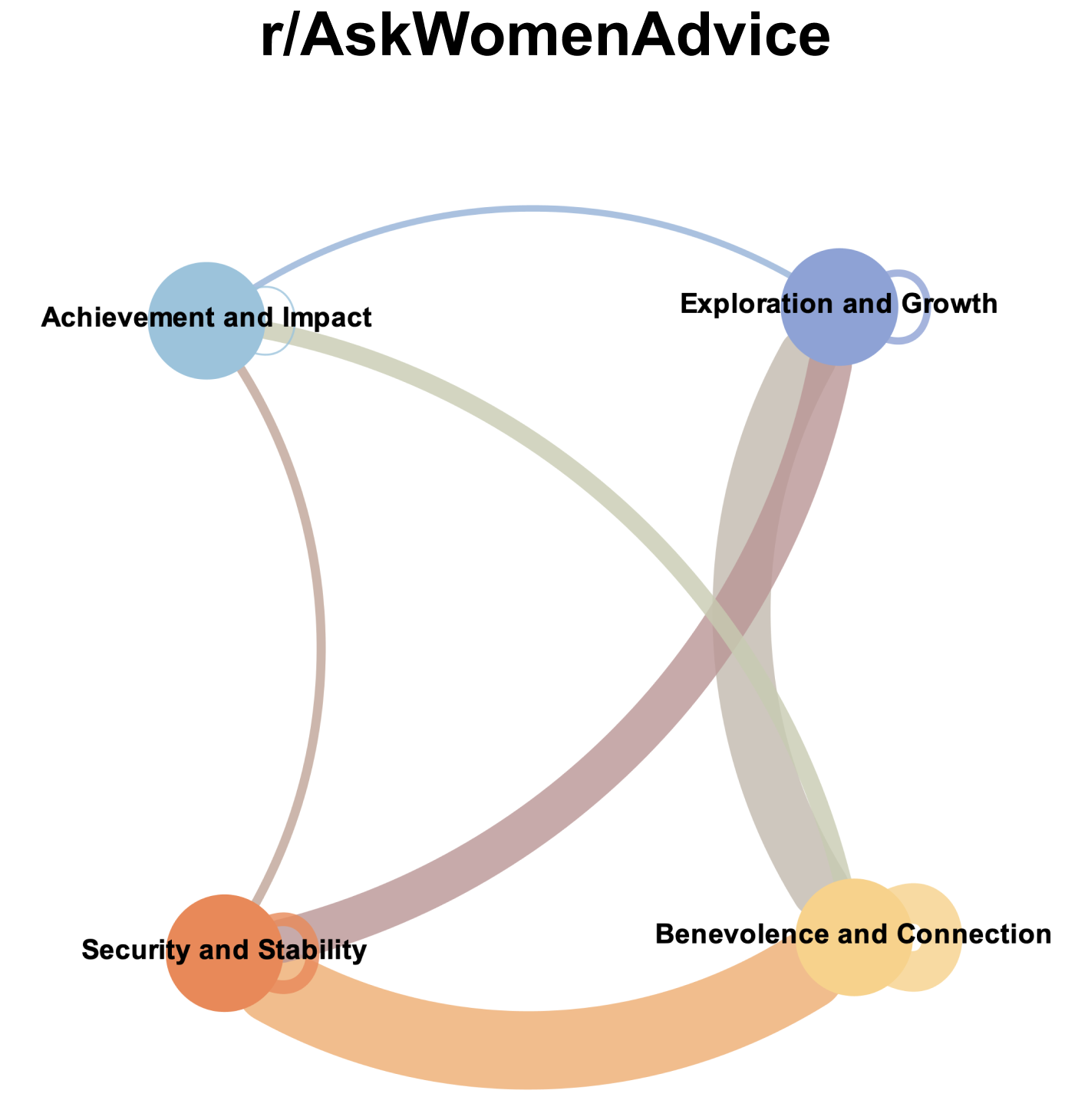}
    \vspace{2pt}
\end{minipage}

\vspace{5pt}

\begin{minipage}{0.48\columnwidth}
    \centering
    \includegraphics[width=0.85\linewidth]{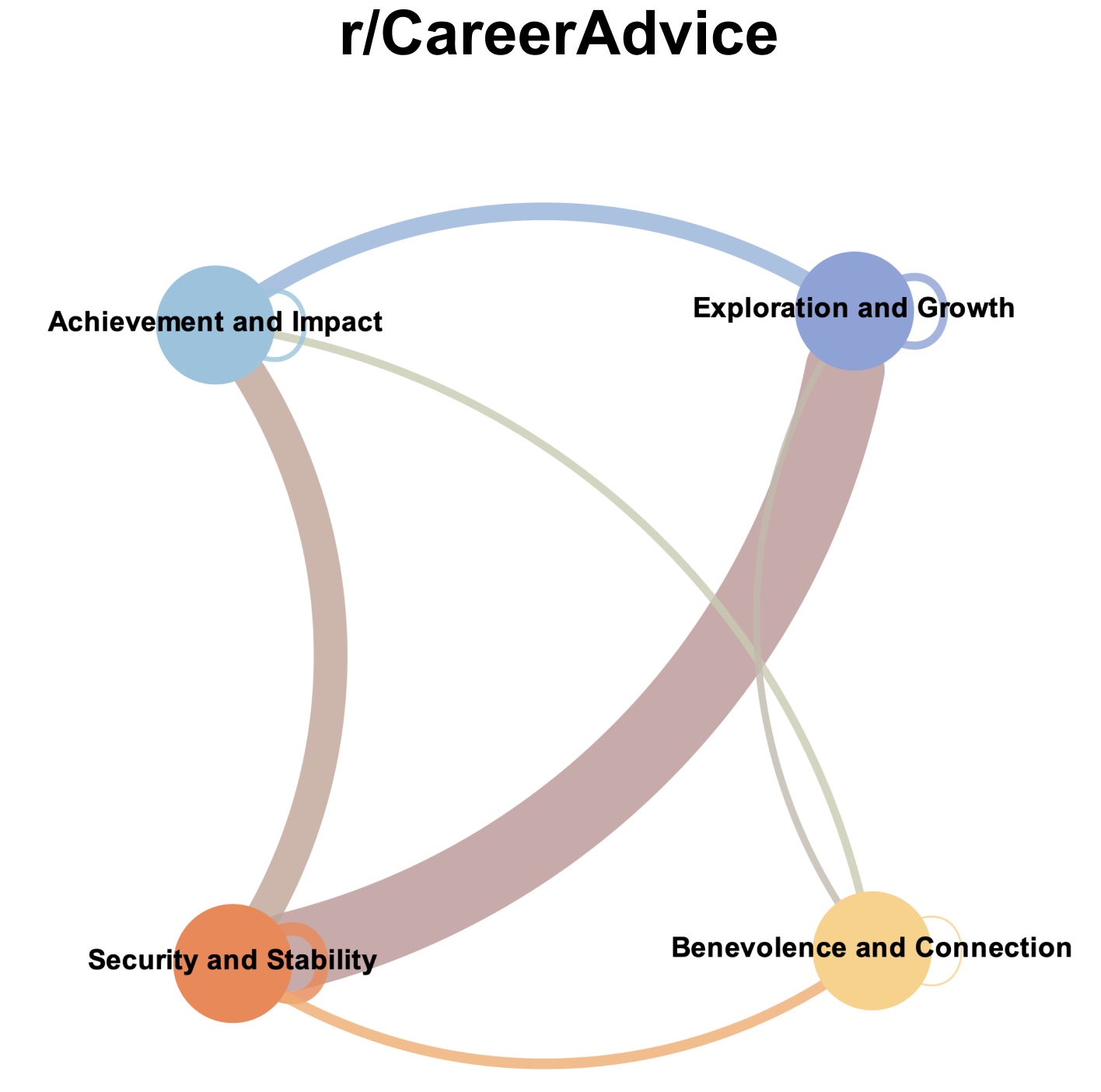}
    \vspace{2pt}
\end{minipage}
\hfill
\begin{minipage}{0.48\columnwidth}
    \centering
    \includegraphics[width=0.85\linewidth]{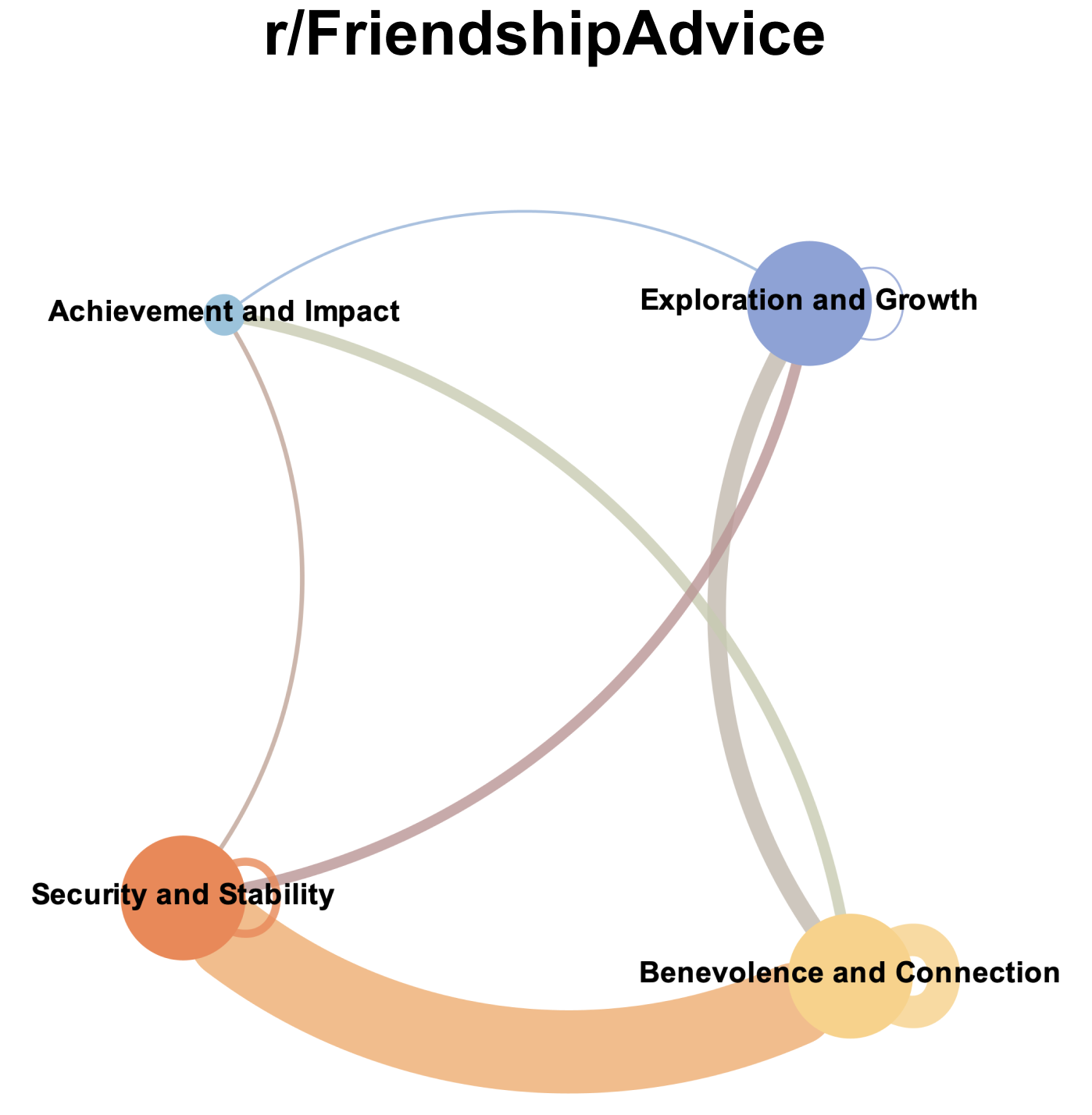}
    \vspace{2pt}
\end{minipage}

\caption{Value co-occurrence networks at Level$_3$ across four advice-oriented subreddits: r/AskMenAdvice, r/AskWomenAdvice, r/CareerAdvice, and r/FriendshipAdvice. Nodes represent individual values and edges indicate co-occurrence frequency within the same dilemma, with thicker edges reflecting stronger co-occurrence.}
\label{fig:value_networks_3}
\end{figure}

\clearpage
\subsection{Value Co-occurrence Networks at Level$_1$}

\begin{figure}[h]
\centering

\begin{minipage}{0.48\columnwidth}
    \centering
    \includegraphics[width=0.85\linewidth]{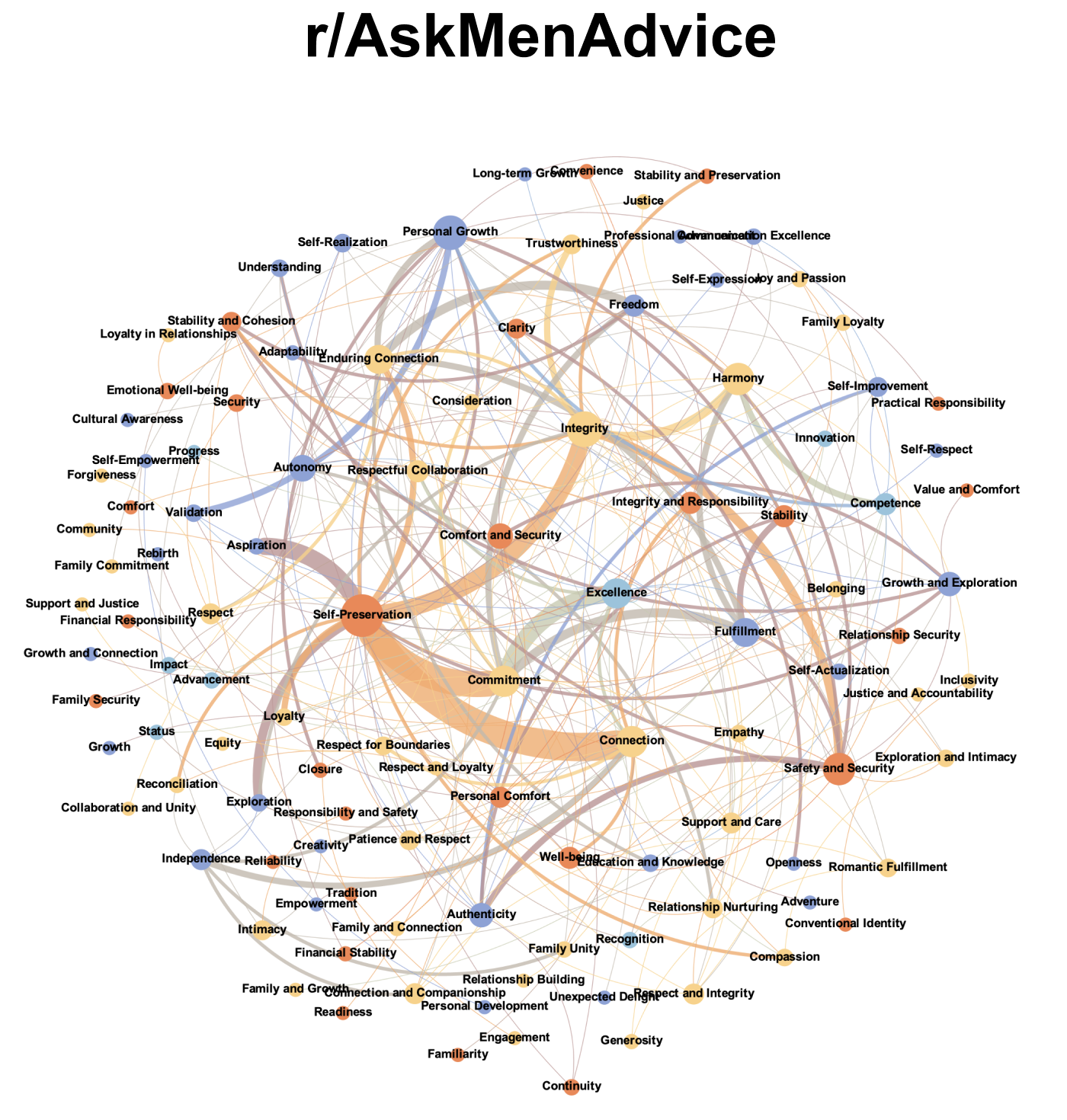}
    \vspace{2pt}
\end{minipage}
\hfill
\begin{minipage}{0.48\columnwidth}
    \centering
    \includegraphics[width=0.85\linewidth]{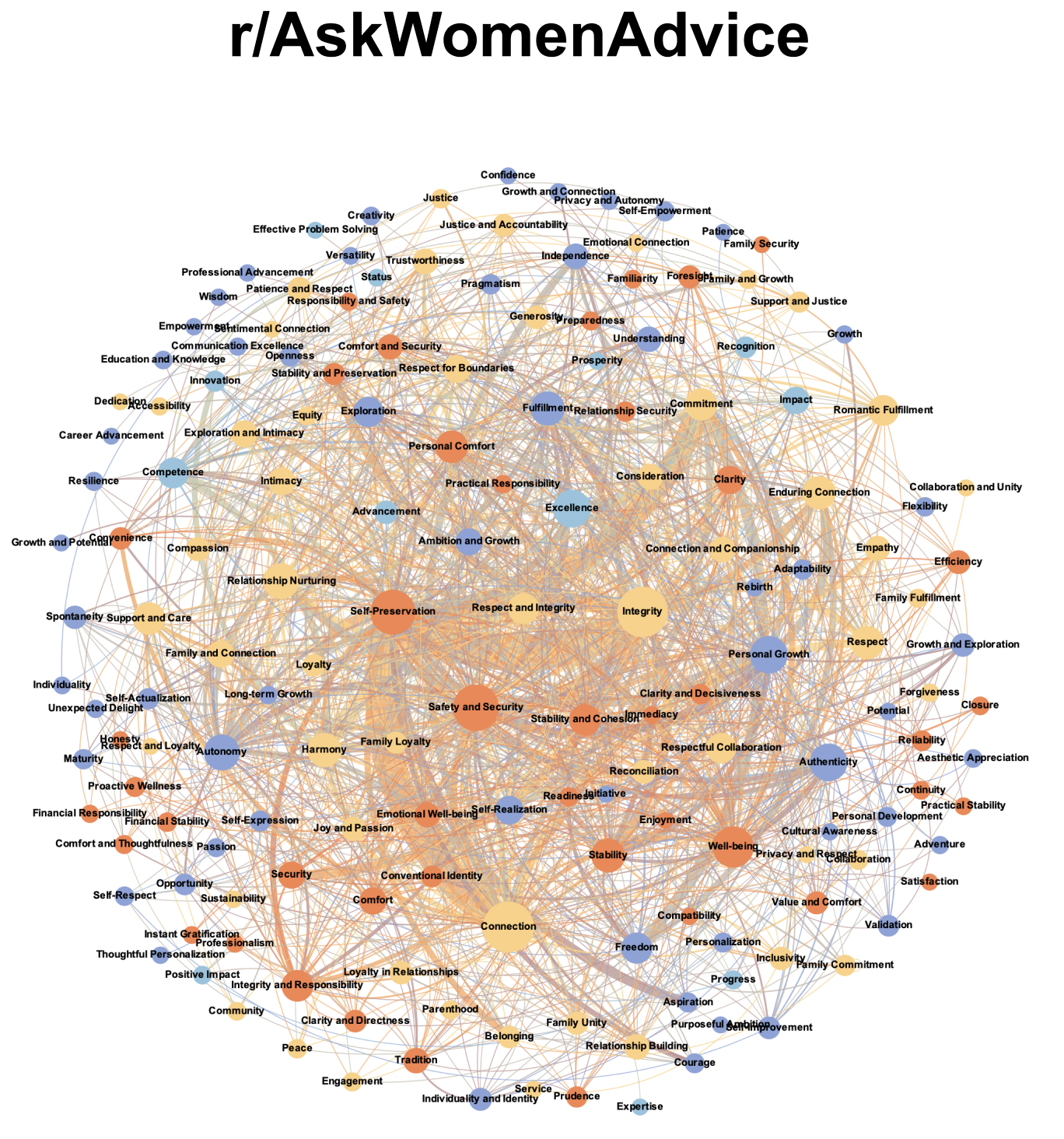}
    \vspace{2pt}
\end{minipage}

\vspace{5pt}

\begin{minipage}{0.48\columnwidth}
    \centering
    \includegraphics[width=0.85\linewidth]{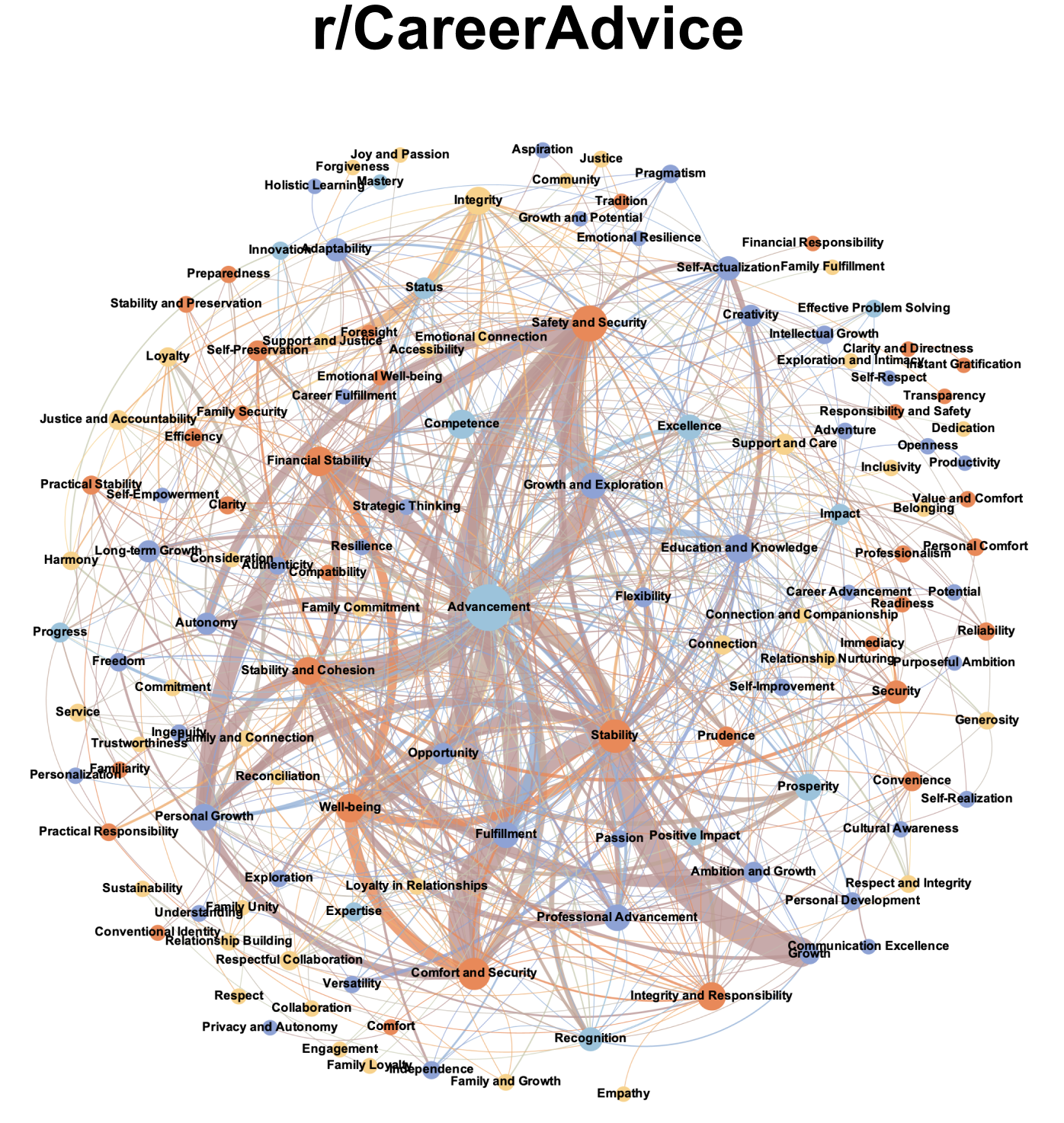}
    \vspace{2pt}
\end{minipage}
\hfill
\begin{minipage}{0.48\columnwidth}
    \centering
    \includegraphics[width=0.85\linewidth]{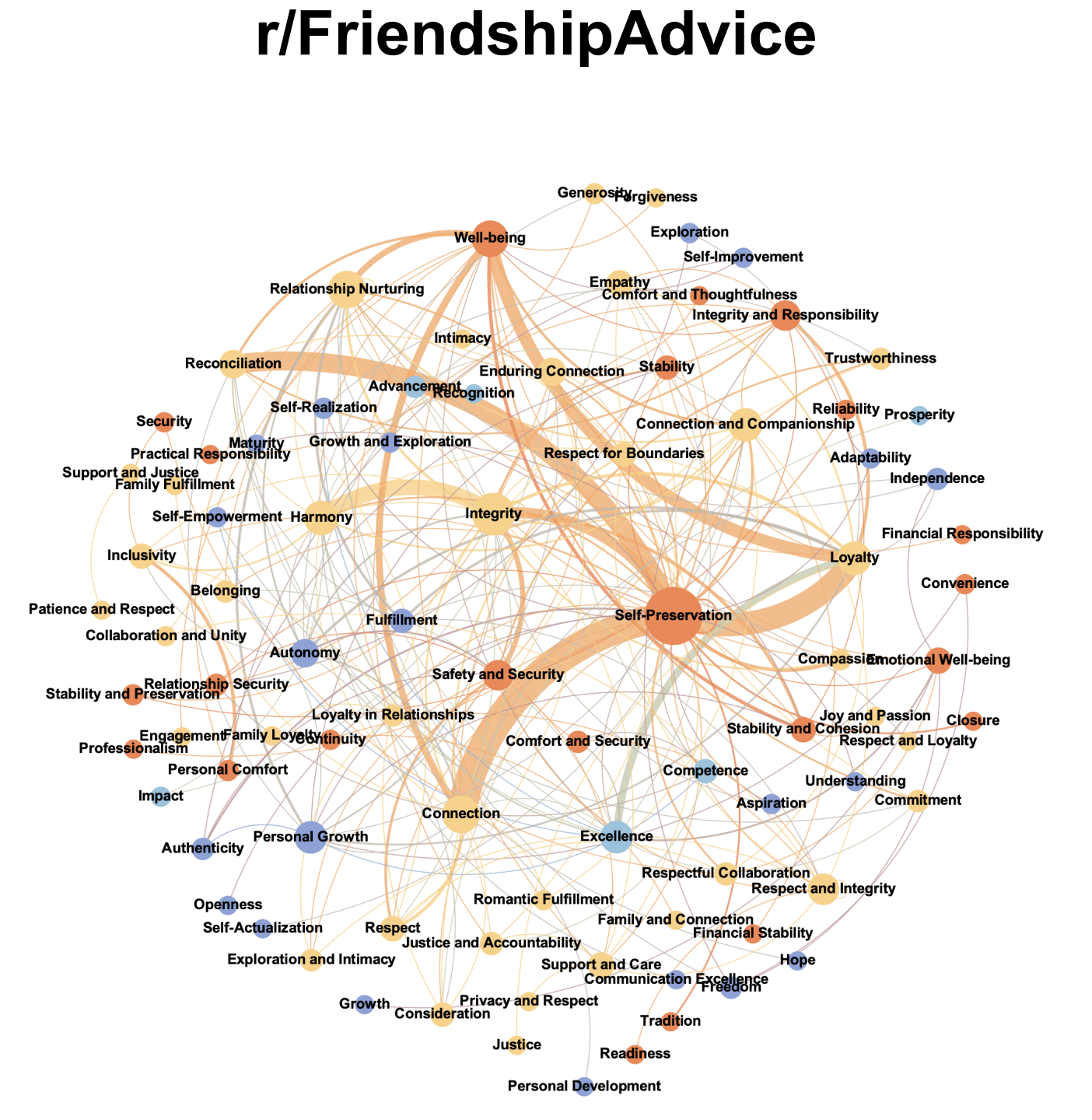}
    \vspace{2pt}
\end{minipage}

\caption{Value co-occurrence networks at Level$_1$ across four advice-oriented subreddits: r/AskMenAdvice, r/AskWomenAdvice, r/CareerAdvice, and r/FriendshipAdvice. Nodes represent individual values and edges indicate co-occurrence frequency within the same dilemma, with thicker edges reflecting stronger co-occurrence. Node colors correspond to the top-level values derived from our hierarchical framework: \textit{Security \& Stability} (orange), \textit{Benevolence \& Connection} (yellow), \textit{Exploration \& Growth} (blue), and \textit{Achievement \& Impact} (light blue). }
\label{fig:value_networks_1}
\end{figure}

\clearpage

\section{Full Results of Significance Tests}
\subsection{Overall Value Preferences of LLMs}
\label{sec: Overall Value}

\vspace*{-\baselineskip}
\vspace{6pt}

\begin{table}[h]
\centering
\scriptsize
\caption{Bootstrapped pairwise comparisons of overall value preferences across models (mean differences). B: Benevolence \& Connection, A: Achievement \& Impact, S: Security \& Stability, E: Exploration \& Growth. Significance levels: $^{*}p<0.05$, $^{**}p<0.01$, $^{***}p<0.001$. Statistically significant cells are highlighted in red.}
\setlength{\tabcolsep}{6pt}
\renewcommand{\arraystretch}{1.1}
\begin{tabular}{@{}lccc@{}}
\toprule
Value Trade-off & GPT-4o & DeepSeek-V3.2-Exp & Gemini-2.5-Flash \\
\midrule
B--A & \cellcolor{red!20}{-0.075***} & \cellcolor{red!20}{-0.145***} & \cellcolor{red!20}{-0.151***} \\
B--S & \cellcolor{red!20}{-0.080***} & \cellcolor{red!20}{-0.074***} & \cellcolor{red!20}{-0.137***} \\
B--E & \cellcolor{red!20}{-0.125***} & \cellcolor{red!20}{-0.159***} & \cellcolor{red!20}{-0.199***} \\
A--S & -0.004 & \cellcolor{red!20}{0.071**} & 0.014 \\
A--E & -0.049 & -0.013 & -0.049 \\
S--E & \cellcolor{red!20}{-0.045*} & \cellcolor{red!20}{-0.085***} & \cellcolor{red!20}{-0.063***} \\
\bottomrule
\end{tabular}
\label{tab:overall_mean_diffs}
\end{table}

\subsection{Within-Subreddit Value Preferences of LLMs}
\label{sec: Within-Subreddit Value}

\vspace*{-\baselineskip}
\vspace{6pt}
\begin{table*}[h]
\centering
\scriptsize
\caption{Bootstrapped pairwise comparisons of value preferences across models and subreddits (mean differences). B: Benevolence \& Connection, A: Achievement \& Impact, S: Security \& Stability, E: Exploration \& Growth. Significance levels: $^{*}p<0.05$, $^{**}p<0.01$, $^{***}p<0.001$. Statistically significant cells are highlighted in red.}
\setlength{\tabcolsep}{8pt}
\renewcommand{\arraystretch}{1.1}
\begin{tabular}{@{}lcccc@{}}
\toprule
 & r/AskMenAdvice & r/AskWomenAdvice & r/CareerAdvice & r/FriendshipAdvice \\
\midrule
\multicolumn{5}{@{}l@{}}{\textbf{GPT-4o}} \\
B-A & -0.125 & -0.042 & \cellcolor{red!20}-0.099* & \cellcolor{red!20}-0.234** \\
B-S & -0.126 & \cellcolor{red!20}-0.067* & -0.046 & -0.055 \\
B-E & \cellcolor{red!20}-0.162* & \cellcolor{red!20}-0.069* & \cellcolor{red!20}-0.181*** & \cellcolor{red!20}-0.337*** \\
A-S & -0.001 & -0.026 & 0.053 & \cellcolor{red!20}0.179*** \\
A-E & -0.037 & -0.027 & \cellcolor{red!20}-0.081* & -0.103 \\
S-E & -0.035 & -0.002 & \cellcolor{red!20}-0.134*** & \cellcolor{red!20}-0.282* \\
\midrule
\multicolumn{5}{@{}l@{}}{\textbf{DeepSeek-V3.2-Exp}} \\
B-A & \cellcolor{red!20}-0.226* & \cellcolor{red!20}-0.099* & \cellcolor{red!20}-0.158** & \cellcolor{red!20}-0.364** \\
B-S & -0.107 & \cellcolor{red!20}-0.084** & -0.054 & \cellcolor{red!20}-0.139* \\
B-E & -0.141 & \cellcolor{red!20}-0.091** & \cellcolor{red!20}-0.205*** & \cellcolor{red!20}-0.345* \\
A-S & 0.119 & 0.014 & \cellcolor{red!20}0.103** & 0.225 \\
A-E & 0.085 & 0.008 & -0.047 & 0.019 \\
S-E & -0.034 & -0.006 & \cellcolor{red!20}-0.150*** & -0.205 \\
\midrule
\multicolumn{5}{@{}l@{}}{\textbf{Gemini-2.5-Flash}} \\
B-A & -0.179 & \cellcolor{red!20}-0.140*** & \cellcolor{red!20}-0.144** & \cellcolor{red!20}-0.347** \\
B-S & -0.127 & \cellcolor{red!20}-0.148*** & -0.063 & \cellcolor{red!20}-0.221*** \\
B-E & \cellcolor{red!20}-0.190* & \cellcolor{red!20}-0.130*** & \cellcolor{red!20}-0.239*** & \cellcolor{red!20}-0.325* \\
A-S & 0.053 & -0.008 & \cellcolor{red!20}0.081* & 0.126 \\
A-E & -0.010 & 0.010 & \cellcolor{red!20}-0.095* & 0.022 \\
S-E & -0.063 & 0.018 & \cellcolor{red!20}-0.176*** & -0.103 \\
\bottomrule
\end{tabular}
\label{tab:pairwise_significance}
\end{table*}

\clearpage

\subsection{Between-Subreddit Value Preferences of LLMs}
\label{sec: Between-Subreddit Value}

\vspace*{-\baselineskip}
\vspace{6pt}

\begin{table}[h]
\centering
\scriptsize
\caption{Bootstrapped pairwise comparisons of subreddits conditional on each value (mean differences). Significance levels: $^{*}p<0.05$, $^{**}p<0.01$, $^{***}p<0.001$. Statistically significant cells are highlighted in red.}
\setlength{\tabcolsep}{6pt}
\renewcommand{\arraystretch}{1.1}
\begin{tabular}{@{}lcccc@{}}
\toprule
 & Benevolence \& Connection & Achievement \& Impact & Security \& Stability & Exploration \& Growth \\
\midrule
\multicolumn{5}{@{}l@{}}{\textbf{GPT-4o}} \\
men--women       & -0.058 & 0.024 & 0.000 & 0.034 \\
men--career      & -0.022 & 0.004 & 0.058 & -0.041 \\
men--friendship  & 0.053 & -0.056 & 0.124 & -0.122 \\
women--career    & 0.037 & -0.020 & \cellcolor{red!20}0.058* & \cellcolor{red!20}-0.075* \\
women--friendship& \cellcolor{red!20}0.112* & -0.080 & \cellcolor{red!20}0.124* & -0.156 \\
career--friendship & 0.075 & -0.060 & 0.066 & -0.081 \\
\midrule
\multicolumn{5}{@{}l@{}}{\textbf{DeepSeek-V3.2-Exp}} \\
men--women       & -0.050 & 0.077 & -0.027 & 0.000 \\
men--career      & -0.014 & 0.054 & 0.038 & -0.078 \\
men--friendship  & 0.094 & -0.044 & 0.061 & -0.110 \\
women--career    & 0.036 & -0.023 & \cellcolor{red!20}0.066* & \cellcolor{red!20}-0.078* \\
women--friendship& \cellcolor{red!20}0.144*** & -0.122 & 0.089 & -0.110 \\
career--friendship & \cellcolor{red!20}0.108* & -0.098 & 0.023 & -0.032 \\
\midrule
\multicolumn{5}{@{}l@{}}{\textbf{Gemini-2.5-Flash}} \\
men--women       & -0.019 & 0.020 & -0.041 & 0.040 \\
men--career      & -0.012 & 0.023 & 0.051 & -0.062 \\
men--friendship  & 0.099 & -0.068 & 0.005 & -0.036 \\
women--career    & 0.007 & 0.003 & \cellcolor{red!20}0.092** & \cellcolor{red!20}-0.102** \\
women--friendship& \cellcolor{red!20}0.119** & -0.088 & 0.046 & -0.076 \\
career--friendship & \cellcolor{red!20}0.112* & -0.091 & -0.047 & 0.026 \\
\bottomrule
\end{tabular}
\label{tab:pairwise_subreddits}
\end{table}

\end{document}